\documentclass[twocolumn,PRL,showpacs,nofootinbib,amsmath,amssymb,superscriptaddress]{revtex4-2}
\pdfoutput=1

\newcommand{\mysection}[1]{\vspace{0.1cm}\noindent\textbf{#1.---}}

\usepackage[utf8]{inputenc}
\usepackage[T1]{fontenc}
\usepackage{microtype}
\usepackage{afterpage,amsmath,amssymb,graphicx,natbib, color}
\usepackage{hyperref}
\usepackage[english]{babel}
\usepackage{blindtext}
\usepackage{csquotes}
\usepackage{enumitem}
\usepackage{graphicx}
\usepackage{hyperref}
\usepackage{mathtools}
\usepackage{comment}
\usepackage{amsmath}
\usepackage{ amssymb }
\usepackage{lipsum} 
\usepackage{float}
\usepackage{cleveref}
\usepackage[caption=false]{subfig}
\usepackage{placeins}
\usepackage{xcolor}
\usepackage{multirow}
\usepackage{float}
\usepackage{makecell}
\usepackage{gensymb}
\usepackage{orcidlink}

\newcommand{\be}{\begin{equation}}
\newcommand{\ee}{\end{equation}}
\newcommand{\bfig}{\begin{figure}}
\newcommand{\efig}{\end{figure}}

\def\ref@jnl#1{{\jnl@style#1}}
\def\aj{\ref@jnl{AJ}}                   
\def\actaa{\ref@jnl{Acta Astron.}}      
\def\araa{\ref@jnl{ARA\&A}}             
\def\apj{\ref@jnl{ApJ}}                 
\def\apjl{\ref@jnl{ApJ}}                
\def\apjs{\ref@jnl{ApJS}}               
\def\ao{\ref@jnl{Appl.~Opt.}}           
\def\apss{\ref@jnl{Ap\&SS}}             
\def\aap{\ref@jnl{A\&A}}                
\def\aapr{\ref@jnl{A\&A~Rev.}}          
\def\aaps{\ref@jnl{A\&AS}}              
\def\azh{\ref@jnl{AZh}}                 
\def\baas{\ref@jnl{BAAS}}               
\def\bac{\ref@jnl{Bull. astr. Inst. Czechosl.}}
\def\caa{\ref@jnl{Chinese Astron. Astrophys.}}
\def\cjaa{\ref@jnl{Chinese J. Astron. Astrophys.}}
\def\icarus{\ref@jnl{Icarus}}           
\def\jcap{\ref@jnl{J. Cosmology Astropart. Phys.}}
\def\jrasc{\ref@jnl{JRASC}}             
\def\memras{\ref@jnl{MmRAS}}            
\def\mnras{\ref@jnl{MNRAS}}             
\def\na{\ref@jnl{New A}}                
\def\nar{\ref@jnl{New A Rev.}}          
\def\pra{\ref@jnl{Phys.~Rev.~A}}        
\def\prb{\ref@jnl{Phys.~Rev.~B}}        
\def\prc{\ref@jnl{Phys.~Rev.~C}}        
\def\prd{\ref@jnl{Phys.~Rev.~D}}        
\def\pre{\ref@jnl{Phys.~Rev.~E}}        
\def\prl{\ref@jnl{Phys.~Rev.~Lett.}}    
\def\pasa{\ref@jnl{PASA}}               
\def\pasp{\ref@jnl{PASP}}               
\def\pasj{\ref@jnl{PASJ}}               
\def\rmxaa{\ref@jnl{Rev. Mexicana Astron. Astrofis.}}%
\def\qjras{\ref@jnl{QJRAS}}             
\def\skytel{\ref@jnl{S\&T}}             
\def\solphys{\ref@jnl{Sol.~Phys.}}      
\def\sovast{\ref@jnl{Soviet~Ast.}}      
\def\ssr{\ref@jnl{Space~Sci.~Rev.}}     
\def\zap{\ref@jnl{ZAp}}                 
\def\nat{\ref@jnl{Nature}}              
\def\iaucirc{\ref@jnl{IAU~Circ.}}       
\def\aplett{\ref@jnl{Astrophys.~Lett.}} 
\def\apspr{\ref@jnl{Astrophys.~Space~Phys.~Res.}}
\def\bain{\ref@jnl{Bull.~Astron.~Inst.~Netherlands}} 
\def\fcp{\ref@jnl{Fund.~Cosmic~Phys.}}  
\def\gca{\ref@jnl{Geochim.~Cosmochim.~Acta}}   
\def\grl{\ref@jnl{Geophys.~Res.~Lett.}} 
\def\jcp{\ref@jnl{J.~Chem.~Phys.}}      
\def\jgr{\ref@jnl{J.~Geophys.~Res.}}    
\def\jqsrt{\ref@jnl{J.~Quant.~Spec.~Radiat.~Transf.}}
\def\memsai{\ref@jnl{Mem.~Soc.~Astron.~Italiana}}
\def\nphysa{\ref@jnl{Nucl.~Phys.~A}}   
\def\physrep{\ref@jnl{Phys.~Rep.}}   
\def\physscr{\ref@jnl{Phys.~Scr}}   
\def\planss{\ref@jnl{Planet.~Space~Sci.}}   
\def\procspie{\ref@jnl{Proc.~SPIE}}   

\begin{document}

\title{A Measurement of CO(3--2) Line Emission from eBOSS Galaxies at $z\sim 0.5$ using Planck Data}

\author{Anirban Roy\,\orcidlink{0000-0001-5729-0246}}
\affiliation{Department of Physics, New York University, 726 Broadway, New York, NY, 10003, USA}  
\affiliation{Center for Computational Astrophysics, Flatiron Institute, New York, NY 10010, USA}
\affiliation{Department of Astronomy, Cornell University, Ithaca, NY 14853, USA}

\author{Nicholas Battaglia\,\orcidlink{0000-0001-5846-0411}}
\affiliation{Department of Astronomy, Cornell University, Ithaca, NY 14853, USA}

\author{Anthony R. Pullen}
\affiliation{Department of Physics, New York University, 726 Broadway, New York, NY, 10003, USA} 
\affiliation{Center for Computational Astrophysics, Flatiron Institute, New York, NY 10010, USA}

\begin{abstract}
Line intensity mapping (LIM) is a novel observational technique in astrophysics that utilizes the integrated emission from multiple atomic and molecular transition lines from galaxies to probe the complex physics of galaxy formation and evolution, as well as the large-scale structure of the universe. Modeling multiple line luminosities of galaxies with varying masses or their host halo masses poses significant uncertainty due to the lack of observational data across a wide redshift range and the intricate nature of astrophysical processes, making them challenging to model analytically or in simulations. While future experiments aim to measure multiple line intensities up to $z\sim 8$ across a wide volume using tomographic methods, we leverage publicly available datasets from the CMB experiment Planck and the galaxy survey eBOSS to constrain the CO(3-2) emission from galaxies. 
We correlate galaxies from eBOSS data onto the full-sky CO(2-1) map produced by Planck and report the first measurement of the average CO(3-2) intensity, $I_{CO} = 45.7 \pm 14.2\, \mathrm{Jy/sr}$ at $z\sim 0.5$ with $3.2\sigma$ confidence. Our findings demonstrate that stacking methods are already viable with existing observations from CMB experiments and galaxy surveys, and are complementary to traditional LIM experiments.
\end{abstract}

\maketitle

\mysection{Introduction}\label{sec:introduction}
Multi-probe cosmology integrates a diverse range of datasets to study the properties and evolution of the universe, including those from cosmic microwave background (CMB) radiation, large-scale structure (LSS) surveys, supernovae, and galaxy clusters \cite{DES-multiprobe, Moresco2022}. Line intensity mapping (LIM) emerges as a promising addition to this multifaceted approach, offering unique insights into galaxy evolution and the large-scale distribution of matter in the universe  \cite{Kovetz2017LIM_report, Kovetz2019-LIM, Schaan:2021gzb}. The ongoing and future LIM experiments aim to map spatial intensity fluctuations of specific spectral lines, including rotational level transitions of CO(1-0) to CO(13-12), [CII], [OIII], H$\alpha$, H$\beta$, 21\,cm and others, based on their brightness feature and frequency overlaps \cite{Time-science-2014, SPHEREx-science-paper2018, CONCERTO-science-2020, EXCLAIM-2020, CCAT-prime2021, COmap-science-2021}. These observations across a broad redshift range up to z $\sim 8$ could provide valuable understanding into the formation of the first stars during the epoch of reionization, cosmic star formation history, metal abundances in galaxies and cosmological information such as the growth of structures, expansion history of the Universe and baryon acoustic oscillations. While emissions of different J-level CO and [CII] from individual galaxies are detected through high-resolution target observations \cite{Greve2014, genzel2015combined, schaerer2020alpine}, the next frontier lies in detecting the spatial fluctuations of these lines which is crucial for extracting both astrophysical and cosmological information \cite{silva2021mapping, padmanabhan2021multi}.

One main challenge arises from modeling the multiple line intensities for different types of galaxies across varying masses and environments, from a few hundred million years after the Big Bang to the present day \cite{Suginohara1998, Righi2008b, Carilli2011, Fonseca2016, Gong2017, Chung2018CII, PadmanabhanCO, Padmanabhan_CII, Dumitru2018, Chung2018CO, Kannan:2021ucy, Murmu:2021quo, Karoumpis2021, limfast2, garcia2023texttt}. Quantifying the various model uncertainties is pivotal, as these could alter the interpretation of the underlying physics of galaxy formation and evolution once the LIM observations are conducted \cite{Murmu:2021ljb, limpy-roy}. Therefore, a promising approach is to consolidate all probes related to galaxy formation and evolution that complement LIM into one framework and perform a joint analysis to minimize these uncertainties. Moreover, precise LIM observations are important for understanding foreground contamination in CMB experiments, particularly for the measurement of secondary anisotropies such as thermal and kinetic Sunyaev-Zeldovich effects (tSZ and kSZ), lensing, and primordial $B$-mode signals, and this will aid in mitigating such contamination for more accurate cosmological parameter estimation \cite{Righi2008b, puglisi:2017eqj, maniyar2023extragalactic}.

\begin{figure*}[t]
\includegraphics[width=\textwidth]{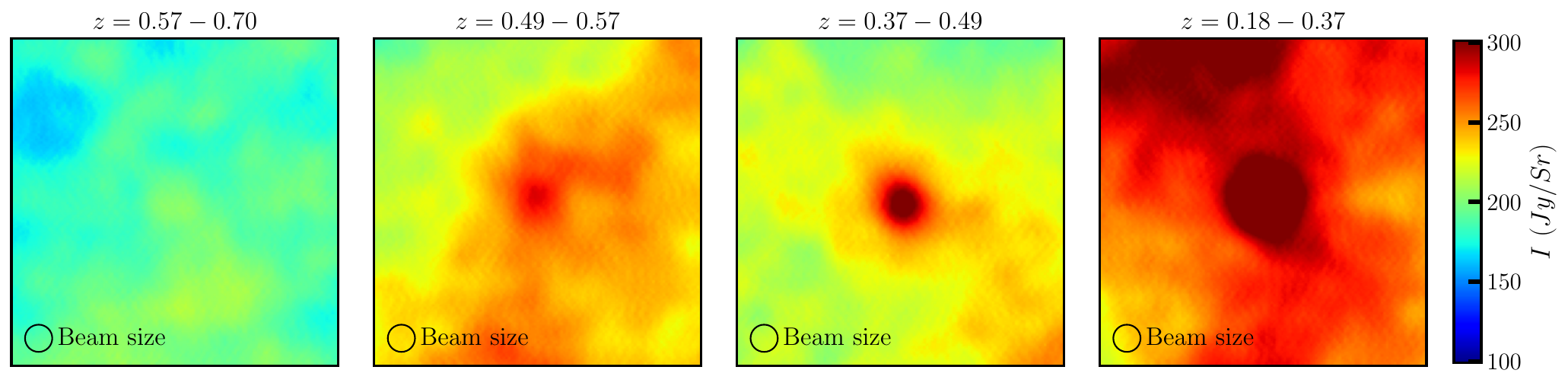} 
\caption{
The panels depict intensity images generated by galaxies from four distinct redshift bins, stacked onto the CO(2-1) Type 2 map derived from Planck data. The panel corresponding to the redshift bin 0.49 to 0.57, aligned with Planck's frequency band of 219.3 to 230.8 GHz, is anticipated to exhibit CO(3-2) emissions. Conversely, signals observed in the other panels reflect the frequency-dependent residual foreground contamination. The FWHM of the beam of the map is assumed to be $4'.99$ \citep{Planck-CO-2013}.}
\label{fig:stacked_maps}
\end{figure*}

The search for [CII] emission at $z\sim2.6$ has been conducted through the cross-correlation of Planck's 353, 545, and 857 GHz data with BOSS quasars and CMASS galaxies \cite{Pullen-CII-2018}. This analysis revealed an excess signal after jointly fitting the [CII] intensity, CIB and tSZ model. The upper limit of CO(1-0) emission from redshifts z = 2.4 to 3.4 has been estimated by cross-correlating eBOSS quasars with the first-year intensity observations from COMAP \cite{Comap-stacking}. In this letter, we investigate the potential of employing a novel stacking method to extract higher-order J-level CO transitions from the CO maps with lower J-level. Drawing from the extensive use of stacking methods in detecting and characterizing galaxy clusters on Planck's tSZ maps \cite{greco2015stacked, aghanim2016planck-tsz-stacking, hill2018two-stacking,  Hall-tsz-2019}, we apply this approach to existing component separated maps from Planck. Specifically, we utilize the full-sky CO(2-1) map generated by Planck to identify the contributions of CO(3-2) from galaxies detected by eBOSS at $z\sim 0.5$ \cite{Planck-CO-2013, eboss-technical, eboss-overview}. The stacking of sources, such as galaxies in a map for our case relies on minimizing the random noise properties while cumulatively adding the weaker signals from each galaxy. This process increases the signal-to-noise ratio (SNR) for the detection of the stacked image, enhancing its detectability and then, inferring the average properties of sources in the stacked image. 

This letter is structured as follows: first, we introduce the methodology used to correlate the galaxies observed by eBOSS with the Planck CO(2-1) map. Next, we characterize potential contamination from other sources, such as residual cosmic infrared background (CIB) and instrumental noise, and interpret the stacked CO(3-2) intensity maps. Finally, we discuss the significance of these results, highlighting the importance of these methods and their potential implications for the current and next generation of LIM experiments. In this letter, we adopt the cosmological model of a flat $\Lambda$CDM universe, as established by the cosmological parameters derived from the Planck TT, TE, EE+lowE+lensing results \citep{P18:main}.

\mysection{Cross-correlation Methodology}
Planck produced CO(1-0) and CO(2-1) maps at the rest-frame frequencies of these lines, i.e., 115 GHz and 230 GHz, despite lacking exact frequency channels to observe the sky at these frequencies \cite{Planck-HFI-spectral-response}. This discrepancy was overcome due to the frequency mismatch of Planck's detectors, which extended their coverage to the rest-frame frequencies of these two lines. Consequently, the Planck satellite was able to map these line emissions with oversampling frequency bands of 109.6 – 115.4 GHz for CO(1-0) and 219.3 – 230.8 GHz for CO(2-1) \cite{Planck-CO-2013}. Line emissions from galaxies situated at redshift $z$ to $z+\Delta z$, corresponding to the frequency bandwidth $\nu$ to $\nu + \Delta \nu$, are expected to be redshifted to Planck's detectors. Additionally, higher-order CO transition lines from galaxies at earlier cosmic times are also redshifted to the same frequency channels. Therefore, Planck CO maps not only comprise prominent CO(1-0) and CO(2-1) lines but also include the subdominant contributions from higher-order CO lines. For instance, CO(3-2) emissions from redshift 0.5724 to 0.4940 are redshifted to Planck's oversampling frequency 219.3 – 230 GHz, and CO(2-1) from redshift 1.1 to 0.99 are redshifted to Planck's CO(1-0) map. With this premise, we decide to look for the CO(3-2) signal from Planck's CO(2-1) maps.

Planck data yield two distinct map types: Type 1 and Type 2 \cite{Planck-CO-2013}. Type 1 maps, derived from spectral transitions of bolometer pairs within a single frequency channel, do not capture intensity contamination from other frequency channels. In contrast, Type 2 maps integrate intensity responses across multiple frequency channels, facilitating the separation of foreground emissions like CMB, free-free, and dust emissions from CO emissions. Comparison of the standard deviations at high latitudes reveals that the uncertainty in Type 2 maps is approximately four times smaller than that in Type 1 maps \cite{Planck-CO-2013}. Consequently, unless specified otherwise, we use CO(2-1) Type 2 maps for our stacking analysis aimed at detecting CO(3-2) signals at redshift $z\sim 0.5$.

Within the redshift range from $z=0.57-0.49$, we utilize the galaxy catalogs of eBOSS to ensure alignment of CO(3-2) emission from these galaxies with the exact frequency bandwidth used in constructing Planck's CO(2-1) map. A total of $427,323$ galaxies within this redshift bin are identified for stacking analysis. Using the Healpy package \cite{healpix}, we identify the positions of individual galaxies on the CO(2-1) map and extract circular regions with a $1.5\degree$ radius, centered on each galaxy's right ascension (RA) and declination (DEC) coordinates. We select this radius to ensure that the resulting map remains within a square area enclosed by the circle, with a side length of $2.12\degree$. We divided this square region into 125 pixels, ensuring that each pixel corresponds to an approximate size of $\sim 1'$. We iterate this process for all the galaxies at this redshift bin and finally, we stack all those cutout maps and make a single stacked image that captures the signature of CO(3-2) emissions from all the galaxies. 

During the stacking analysis, the presence of any correlated signal, rather than random noise, is anticipated to appear at the centre of the image, given that emissions primarily originate from galaxies. However, such observations do not definitively ascertain whether the signal originates solely from CO(3-2) emission or from residual foregrounds correlated with the galaxy positions. To explore this, we employ a stacking approach on galaxies from adjacent redshift bins. This allows us to ascertain if CO(3-2) emission is absent in stacked images composed of galaxies from other redshifts. To ensure consistency in our analysis, an equal number of galaxies are selected from subsequent redshift bins from which we do not expect CO(3-2) emission.

Subsequently, we identify adjacent redshift bins to $z_i$ and $z_i + \Delta z_i$ in a manner ensuring that each bin contains the same number of galaxies. Similar to the redshift bin of CO(3-2) as described previously, we conduct stacking analysis on these adjacent bins using the Planck CO(2-1) Type 2 map.

We present the result of the stacking analysis at four redshifts in Figure \ref{fig:stacked_maps}. We do not find any significant signal in the stacked image of galaxies situated between redshifts 0.57 and 0.70, as expected due to the absence of CO(3-2) emission at these redshifts. The emissions in the redshift range of 0.49 to 0.57 indicate where CO(3-2) is expected to appear, shown in the second panel from the left. Furthermore, as we reach to the low redshift bins, we do not see the signal disappear which would be expected if the intensity signal came \textit{purely} from CO(3-2) emission. In contrast, we see the mean amplitude of the signal at the center of the stacked map increases significantly, confirming galaxies present in the maps, and CO(3-2) signal adds up with the residual signal only on the redshift bin 0.49 to 0.57. It is evident that the size of the central region increases due to the fact that the angular size of galaxies increases at the low redshifts than the high redshift, hence the intensity distribution spreads over more pixels at low redshift than at high redshifts.  

We assess the intensity of galaxies by employing an aperture photometry technique facilitated by the Astropy package \cite{Astropy-5}. Using a central circle radius of 10 arcminutes, we capture the primary intensity concentration attributed to the stacked galaxies. Initially, we determine the intensity of this concentration and subsequently compute the variance of the intensity measurement relative to the background intensity distribution. To accomplish this, we randomly select 100 circles outside the central zone (10 arcminutes) and compute the intensity variance of these outer regions. This approach allows us to quantify both the intensity of emissions originating from galaxies and the errors stemming from random background noise. 

We conduct several tests to ensure the efficacy of the stacking analysis on our target galaxies and to ascertain the appropriateness of signals from various redshift bins for a model encompassing CO(3-2) signal and residual foregrounds. For the initial test, we use the same number of galaxies corresponding to the redshift bin for the emission of CO(3-2) and randomly disperse them across the sky. Due to variations in their RA and DEC values, galaxies in the eBOSS catalog do not match this distribution. Afterwards, we perform the stacking analysis on these randomly distributed galaxies projected to Planck's CO(2-1) Type 2 map as described previously. Our results reveal a lack of observable structures at the center of the stacked map, but trace the presence of homogeneous random noise across the entire map. Moreover, the overall amplitude is approximately three times smaller than the intensity observed at the core of the stacked eBOSS galaxies on the CO(2-1) map. This outcome confirms that the signal observed in Figure \ref{fig:stacked_maps} indeed originates from the eBOSS galaxies within the selected redshift bins.

\begin{figure}
    \centering    \includegraphics[width=0.5\textwidth]{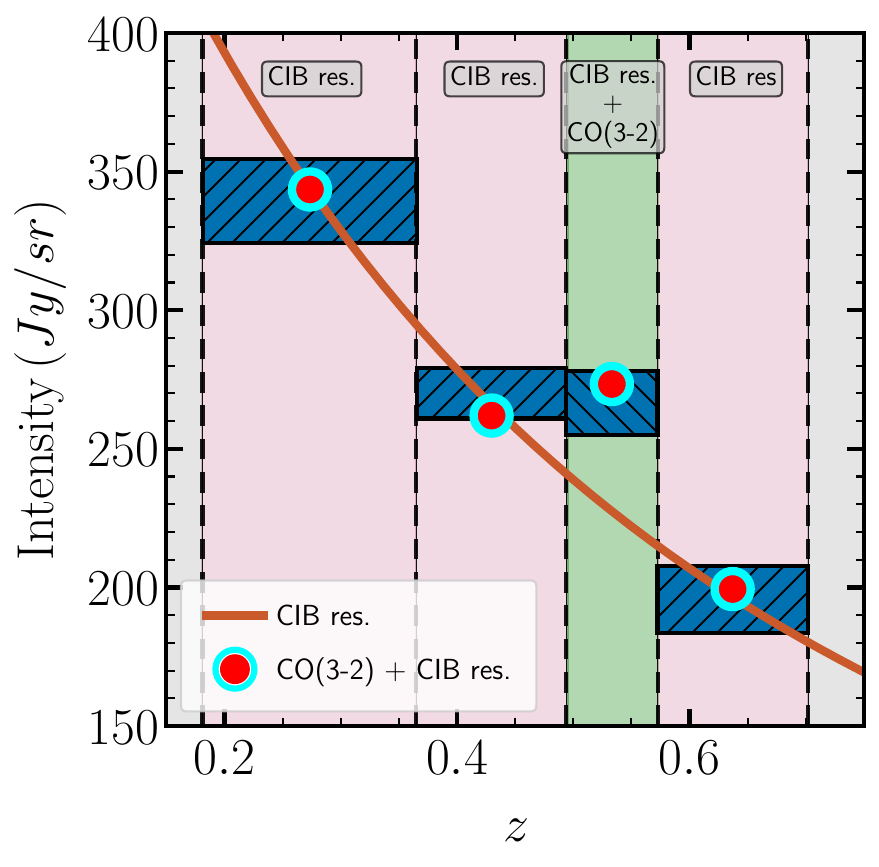}
    \caption{
We display the estimated intensities along with their respective errors with boxes at four redshift bins. The red line represents the residual CIB spectrum calculated using the best-fit parameters. The shaded green region indicates the redshift range where we anticipate contributions from CO(3-2), while the shaded red regions indicate the residual CIB emissions at those redshift bins.}
    \label{fig:co32-detection}
\end{figure}

\mysection{Interpretation}
We proceed to quantify the remaining foreground in the stacked maps of CO(3-2) across four distinct redshift bins, in order to understand the emission we are seeing in these bins. Calculating the intensities of the stacked image across four adjacent redshift bins provides insights into how residual CIB varies with observed frequency (or redshift). It is anticipated that the CIB spectrum should closely resemble a blackbody spectrum; however, the amplitude is expected to be reduced due to the component separation process \cite{Planck-CO-2013}. We model the residual CIB emission using a power-law parameterization as $I_{\rm res}(z) = I^{0}_{res} \times (1+z)^{-\alpha}$, where $I^{0}_{res}$ is the amplitude of residual foreground at $z\sim 0$ and $\alpha$ is the power-law index. 
\begin{figure}[h]
\includegraphics[width=0.49\textwidth]{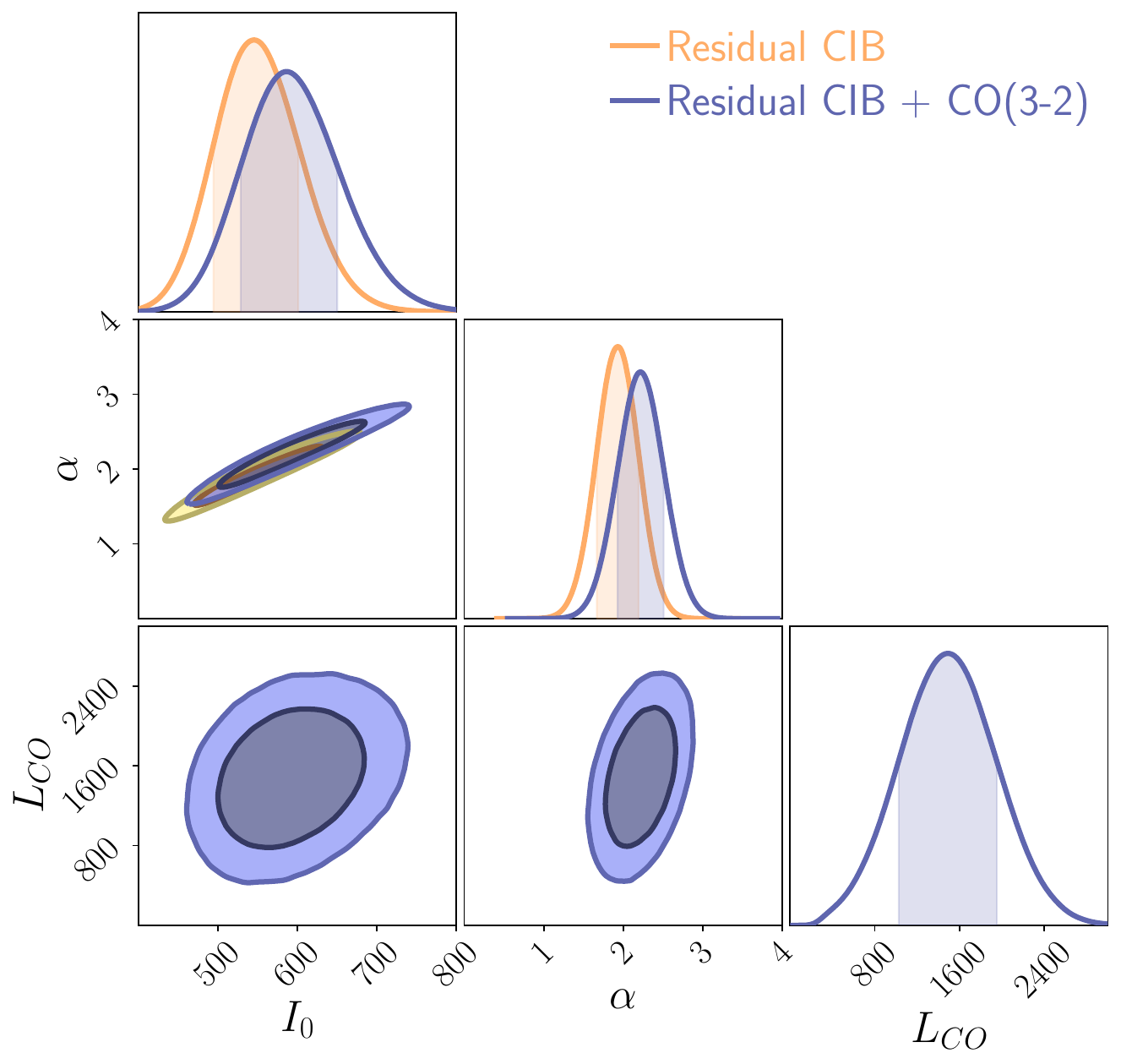} 
\caption{The constraints on parameters for residual CIB and average CO(3-2) luminosity at $z\sim 0.5$ estimated from the stacking analysis with eBOSS galaxies onto the Planck's CO(2-1) map.}
\label{fig:triangle_plot}
\end{figure}

To ascertain the validity of this assumption for the foreground modelling, we performed stacking analysis using an equal number of galaxies across four redshift bins with a CO(1-0) map, as opposed to using CO(2-1) map as previously described. Due to a frequency mismatch, we do not expect any CO emissions for the same redshift bins to be present in the CO(1-0) map. The results, detailed in Figure \ref{fig:sub_CIB} in the supplementary section, reveal a diminished signal, with no prominent structures discernible at the center of the stacked image. However, upon fitting the intensity data points for this scenario using a model considering only residual CIB, we derived $ I^{0}_{res}= 133^{+14}_{-12}\,(Jy/sr)$ and $\alpha = 0.52^{+0.25}_{-0.24}$. In this instance, the chi-square value is 0.55 with 2 degrees of freedom, yielding a probability-to-exceed (PTE) value of 0.76. As this model can perfectly fit the residual foregrounds at the 115 GHz map, we apply the same foreground model to probe the CO(3-2) signal in the 217\, GHz channel. 

\begin{table}[h!!]
\centering
\renewcommand{\arraystretch}{1.25} 
\begin{tabular}{p{0.3\linewidth}p{0.3\linewidth}p{0.3\linewidth}}
\hline
\hline
Parameter & without CO & with CO \\
\hline
$I_0$ & $545^{+56}_{-51}$ & $587^{+63}_{-59}$ \\
$\alpha$ & $1.94^{+0.29}_{-0.27}$ & $2.22 \pm 0.29$ \\
$L_{\rm CO(3-2)}$ & N/A & $1495^{+460}_{-465}$ \\
$\chi^2$ & 7.95 & 1.31 \\
PTE & 0.018 & 0.25 \\
\hline
\end{tabular}
\caption{Best fit parameters with 68\% confidence intervals, accompanied by the corresponding values of $\chi^2$ and PTE, obtained from fitting the intensities of stacked profiles at four redshifts. The fits consider contributions from residual CIB alone and residual CIB with CO(3-2) emissions.}
\label{tab:parameter_comparison}
\end{table}

We construct a model that combines both residual CIB and CO(3-2) emissions, expressed as $I_{\rm stack} = I_{\rm res} + I_{\rm CO(3-2)}$. The CO(3-2) emission model is assumed to originate from the specific redshift bin of 0.49-0.57; therefore, we introduce a free parameter $L_{\rm CO(3-2)}$, representing the average CO(3-2) luminosity of galaxies within that bin. This parameter relates to the intensity as $I_{\rm CO(3-2)} = L_{\rm CO(3-2)} \times (c/4\pi \nu_{\rm rest}H(z))$, where $\nu_{\rm rest}$ is the rest-frame frequency of CO(3-2) and $H(z)$ is the Hubble parameter at the redshift of CO(3-2) emission. To measure the average intensity of CO(3-2), we integrate $dn/dz$ of galaxies in redshift bins and compute the average intensity or foreground as $\langle I \rangle= \int_{z_{\rm min}}^{z_{\rm max}} I(z) \frac{dn(z)}{dz} dz/ \int_{z_{\rm min}}^{z_{\rm max}} \frac{dn}{dz} dz $. Here, $I(z)$ is the total intensity at a given redshift, including residual foreground and CO(3-2) signal, and $dn/dz$ is the redshift distribution of eBOSS galaxies. 

In Figure \ref{fig:co32-detection}, we illustrate how the intensity of the stacked images, encompassing CO(3-2) emissions around $z \sim 0.5$ and residual CIB, varies with redshift. As expected, the intensity of the residual CIB decreases at the high redshift showing a power-law trend. We estimate the intensity at the redshift bins 0.18--0.37, 0.37--0.49, and 0.57--0.70 to be $339.6 \pm 15.1$, $270 \pm 9.2$, and $195.5 \pm 12$ Jy/sr, and the intensity in the redshift bin where CO(3-2) contributes to be $266 \pm 11.5$ Jy/sr. We make use of the emcee package \cite{emcee} to fit the parameters of the intensity model to the estimated intensities at four redshift bins. We consider two scenarios: i) fitting the data while considering only residual CIB present in the stacked map, where the parameters are $I^{0}_{res}$ and $\alpha$, and ii) considering both residual CIB and CO at redshift bin $z \sim 0.5$, where we introduce an additional parameter $L_{\rm CO(3-2)}$ and subsequently calculate the intensity of CO(3-2) from luminosity. We show the constraints on parameters through a triangle plot in Figure \ref{fig:triangle_plot} and the best-fit parameters for both cases are reported in Table \ref{tab:parameter_comparison}. When fitting the data with a CIB-only model, the $\chi^2$ becomes 7.95 with 2 degrees of freedom, yielding a PTE value of 0.018. Conversely, fitting the data with residual CIB and CO(3-2) emission reduces the $\chi^2$ value to 1.31, with a corresponding PTE value of 0.25. Therefore, our analysis confirms that the presence of CO(3-2) emission, along with residual foreground, provides a better description of the observed data. 

\begin{figure}[h]
\includegraphics[width=0.49\textwidth]{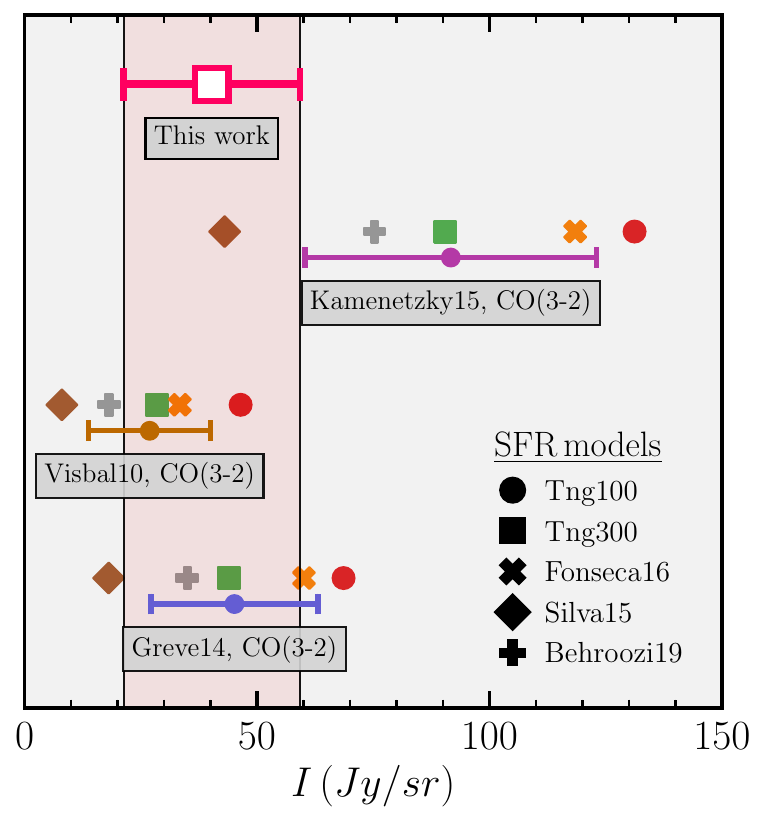} 
\caption{Various CO(3-2) model predictions are compared with the measured CO(3-2) intensity at $z\sim 0.5$. The legend provides descriptions of the SFR models, while annotations within the gray boxes denote the models that convert SFR to CO(3-2) line luminosities. The horizontal bar for a particular line luminosity model is calculated from the variance in intensities for the different SFR models.}
\label{fig:model_comparison}
\end{figure}

This measurement translates to an average CO(3-2) luminosity density of $L_{\rm CO}=1495^{+460}_{-465}\,L_\odot/{Mpc}^3$ at $z\sim 0.5$, with the corresponding intensity constrained to $I_{\rm CO}=45.7 \pm 14.2$ with 68\% confidence. Predicting the intensity of CO(3-2) analytically is challenging due to various factors: i) the relationship between CO(3-2) luminosity and star formation rate (SFR) or halo mass is not precisely constrained and depends on multiple empirical scaling relations, ii) the ratio of CO(3-2) to CO(1-0) or other CO J-level transitions is not firmly established across a wide redshift range, iii) observed CO(3-2) intensity may deviate from estimated intensity due to differences in integrating luminosity over all halo masses versus specific galaxy selection criteria, among other factors. Considering these complexities, we compare our measurement with the estimated values with those in the literature using different SFR models and scaling relations between SFR and CO(3-2) luminosity, Visbal10 \citep{Visbal2010}, Greve14 \citep{Greve2014}, and Kamenetzky15 \citep{Kamenetzky2016}. We estimate the average intensity of CO(3-2) lines for eBOSS galaxies using the formula \(I_{\text{est}}=\frac{c}{4\pi \nu_{\text{rest}} H(z=0.5)} \times \int_{M_{\text{min}}}^{M_{\text{max}}} L(M) \frac{dn}{dM} dM\). Here, \(M_{\text{min}}\) and \(M_{\text{max}}\) represent the minimum and maximum stellar masses of the galaxy samples, respectively. We present our findings in Figure \ref{fig:model_comparison}. Details of the model description and theoretical background are provided in \cite{limpy-roy}. We calculate the average intensity of CO(3-2) using five SFR models: Tng100, Tng300 \citep{Nelson:2018uso, TNG-gal, TNG-gen}, Fonseca16 \citep{Fonseca2016}, Silva15 \citep{Silva:2015}, and Behroozi19 \citep{Behroozi2019} as described in \citep{limpy-roy}. The variance and mean of these intensities corresponding to SFR models are determined, and the error is calculated from the variance. Our measurement is in good agreement with Visbal10 and Greve14, as most of these models fall within the $1\sigma$ uncertainty of our measurement.

\mysection{Conclusion and Outlook}\label{sec:discussion}
In this letter, we investigate the feasibility of employing stacking analysis as a novel method to constrain the average intensity of line emissions. 
We present the first constraints on the luminosity and intensity of CO(3-2) by employing this method on existing data from Planck and eBOSS. We constrained $I_{CO} = 45.7 \pm 14.2\, \mathrm{Jy/sr}$ at $z\sim 0.5$ with $3.2\sigma$ confidence. Comparing our results to three empirical models of CO(3-2) emission we found that they were in agreement with our measurement, but this agreement depended upon the halo mass to star-formation relation chosen. When used on forthcoming line intensity mapping  datasets, we expect a higher SNR yield for similar cross-correlation analyses, and such analyses will have the potential to further our understanding of galaxy evolution \citep{Breysse2022-co, Comap-stacking,Horlaville2024}.
Forthcoming LIM cross-correlation analyses like this will be able to narrow down the predictions for extragalactic CO contamination in broadband CMB observations \citep[e.g.,][]{maniyar2023extragalactic,Kokron24}.

The method discussed in this letter complements the more commonly discussed LIM analysis at the power spectrum level. Analyzing LIM at the power spectrum has the potential to constrain the mean intensity of lines in a tomographic manner, aiding in constraining the average line intensity. However, a degeneracy exists between the average intensity and bias parameters, with ongoing experiments able to place limits on the combined of two, $b \times I_{\nu}$ parameter. In contrast, $I_{\nu}$ can be estimated solely through stacking analysis for a given sample of galaxies, independent of the bias parameter. Therefore, assimilating both probes—stacking and power spectrum analysis—could lead to tighter constraints on the average intensity parameter as well as other astrophysical parameters.

Our work is a first step towards detecting the intensity of a particular line by correlating galaxies with an intensity map and developing the pipeline for data analysis. In the future, an extensive array of galaxy samples will become available from experiments such as DESI, LSST, and Euclid, accompanied by low-noise LIM maps from experiments like EoR-Spec at FYST, COMAP, and CONCERTO, thereby expanding the scientific potential of stacking analysis. Following the reported detection of the stacking of CO(3-2) lines of galaxies in this letter, we plan to present a detailed analysis of robust forecasts for future experiments in our subsequent works.

We thank Dongwoo Chung, Colin Hill, Abhishek Maniyar, Matt McQuinn, and Alex van Engelen for their helpful discussions. AR acknowledges support from NASA under award number 80NSSC18K1014939 during the final stage of this work. NB acknowledges support from CCAT collaboration for this work and additional support from NASA grants 80NSSC18K0695 and 80NSSC22K0410. ARP was supported
by NASA under award numbers 80NSSC18K1014,
NNH17ZDA001N, and 80NSSC22K0666, and by the NSF under award number 2108411.

\bibliographystyle{apsrev}
\bibliography{citation} 

\begin{thebibliography}{59}
\expandafter\ifx\csname natexlab\endcsname\relax\def\natexlab#1{#1}\fi
\expandafter\ifx\csname bibnamefont\endcsname\relax
  \def\bibnamefont#1{#1}\fi
\expandafter\ifx\csname bibfnamefont\endcsname\relax
  \def\bibfnamefont#1{#1}\fi
\expandafter\ifx\csname citenamefont\endcsname\relax
  \def\citenamefont#1{#1}\fi
\expandafter\ifx\csname url\endcsname\relax
  \def\url#1{\texttt{#1}}\fi
\expandafter\ifx\csname urlprefix\endcsname\relax\def\urlprefix{URL }\fi
\providecommand{\bibinfo}[2]{#2}
\providecommand{\eprint}[2][]{\url{#2}}

\bibitem[{\citenamefont{Abbott et~al.}(2019)\citenamefont{Abbott, Alarcon,
  Allam, Andersen, Andrade-Oliveira, Annis, Asorey, Avila, Bacon, Banik
  et~al.}}]{DES-multiprobe}
\bibinfo{author}{\bibfnamefont{T.}~\bibnamefont{Abbott}},
  \bibinfo{author}{\bibfnamefont{A.}~\bibnamefont{Alarcon}},
  \bibinfo{author}{\bibfnamefont{S.}~\bibnamefont{Allam}},
  \bibinfo{author}{\bibfnamefont{P.}~\bibnamefont{Andersen}},
  \bibinfo{author}{\bibfnamefont{F.}~\bibnamefont{Andrade-Oliveira}},
  \bibinfo{author}{\bibfnamefont{J.}~\bibnamefont{Annis}},
  \bibinfo{author}{\bibfnamefont{J.}~\bibnamefont{Asorey}},
  \bibinfo{author}{\bibfnamefont{S.}~\bibnamefont{Avila}},
  \bibinfo{author}{\bibfnamefont{D.}~\bibnamefont{Bacon}},
  \bibinfo{author}{\bibfnamefont{N.}~\bibnamefont{Banik}},
  \bibnamefont{et~al.}, \bibinfo{journal}{Physical review letters}
  \textbf{\bibinfo{volume}{122}}, \bibinfo{pages}{171301}
  (\bibinfo{year}{2019}).

\bibitem[{\citenamefont{{Moresco} et~al.}(2022)\citenamefont{{Moresco},
  {Amati}, {Amendola}, {Birrer}, {Blakeslee}, {Cantiello}, {Cimatti},
  {Darling}, {Della Valle}, {Fishbach} et~al.}}]{Moresco2022}
\bibinfo{author}{\bibfnamefont{M.}~\bibnamefont{{Moresco}}},
  \bibinfo{author}{\bibfnamefont{L.}~\bibnamefont{{Amati}}},
  \bibinfo{author}{\bibfnamefont{L.}~\bibnamefont{{Amendola}}},
  \bibinfo{author}{\bibfnamefont{S.}~\bibnamefont{{Birrer}}},
  \bibinfo{author}{\bibfnamefont{J.~P.} \bibnamefont{{Blakeslee}}},
  \bibinfo{author}{\bibfnamefont{M.}~\bibnamefont{{Cantiello}}},
  \bibinfo{author}{\bibfnamefont{A.}~\bibnamefont{{Cimatti}}},
  \bibinfo{author}{\bibfnamefont{J.}~\bibnamefont{{Darling}}},
  \bibinfo{author}{\bibfnamefont{M.}~\bibnamefont{{Della Valle}}},
  \bibinfo{author}{\bibfnamefont{M.}~\bibnamefont{{Fishbach}}},
  \bibnamefont{et~al.}, \bibinfo{journal}{Living Reviews in Relativity}
  \textbf{\bibinfo{volume}{25}}, \bibinfo{eid}{6} (\bibinfo{year}{2022}),
  \eprint{2201.07241}.

\bibitem[{\citenamefont{Kovetz et~al.}(2017)\citenamefont{Kovetz, Viero, Lidz,
  Newburgh, Rahman, Switzer, Kamionkowski, Aguirre, Alvarez, Bock
  et~al.}}]{Kovetz2017LIM_report}
\bibinfo{author}{\bibfnamefont{E.~D.} \bibnamefont{Kovetz}},
  \bibinfo{author}{\bibfnamefont{M.~P.} \bibnamefont{Viero}},
  \bibinfo{author}{\bibfnamefont{A.}~\bibnamefont{Lidz}},
  \bibinfo{author}{\bibfnamefont{L.}~\bibnamefont{Newburgh}},
  \bibinfo{author}{\bibfnamefont{M.}~\bibnamefont{Rahman}},
  \bibinfo{author}{\bibfnamefont{E.}~\bibnamefont{Switzer}},
  \bibinfo{author}{\bibfnamefont{M.}~\bibnamefont{Kamionkowski}},
  \bibinfo{author}{\bibfnamefont{J.}~\bibnamefont{Aguirre}},
  \bibinfo{author}{\bibfnamefont{M.}~\bibnamefont{Alvarez}},
  \bibinfo{author}{\bibfnamefont{J.}~\bibnamefont{Bock}}, \bibnamefont{et~al.},
  \bibinfo{journal}{arXiv preprint arXiv:1709.09066}  (\bibinfo{year}{2017}).

\bibitem[{\citenamefont{Kovetz et~al.}(2020)}]{Kovetz2019-LIM}
\bibinfo{author}{\bibfnamefont{E.~D.} \bibnamefont{Kovetz}}
  \bibnamefont{et~al.}, \bibinfo{journal}{Bull. Am. Astron. Soc.}
  \textbf{\bibinfo{volume}{51}}, \bibinfo{pages}{101} (\bibinfo{year}{2020}),
  \eprint{1903.04496}.

\bibitem[{\citenamefont{Schaan and White}(2021)}]{Schaan:2021gzb}
\bibinfo{author}{\bibfnamefont{E.}~\bibnamefont{Schaan}} \bibnamefont{and}
  \bibinfo{author}{\bibfnamefont{M.}~\bibnamefont{White}},
  \bibinfo{journal}{JCAP} \textbf{\bibinfo{volume}{05}}, \bibinfo{pages}{068}
  (\bibinfo{year}{2021}), \eprint{2103.01964}.

\bibitem[{\citenamefont{{Crites} et~al.}(2014)\citenamefont{{Crites}, {Bock},
  {Bradford}, {Chang}, {Cooray}, {Duband}, {Gong}, {Hailey-Dunsheath},
  {Hunacek}, {Koch} et~al.}}]{Time-science-2014}
\bibinfo{author}{\bibfnamefont{A.~T.} \bibnamefont{{Crites}}},
  \bibinfo{author}{\bibfnamefont{J.~J.} \bibnamefont{{Bock}}},
  \bibinfo{author}{\bibfnamefont{C.~M.} \bibnamefont{{Bradford}}},
  \bibinfo{author}{\bibfnamefont{T.~C.} \bibnamefont{{Chang}}},
  \bibinfo{author}{\bibfnamefont{A.~R.} \bibnamefont{{Cooray}}},
  \bibinfo{author}{\bibfnamefont{L.}~\bibnamefont{{Duband}}},
  \bibinfo{author}{\bibfnamefont{Y.}~\bibnamefont{{Gong}}},
  \bibinfo{author}{\bibfnamefont{S.}~\bibnamefont{{Hailey-Dunsheath}}},
  \bibinfo{author}{\bibfnamefont{J.}~\bibnamefont{{Hunacek}}},
  \bibinfo{author}{\bibfnamefont{P.~M.} \bibnamefont{{Koch}}},
  \bibnamefont{et~al.}, in \emph{\bibinfo{booktitle}{Millimeter, Submillimeter,
  and Far-Infrared Detectors and Instrumentation for Astronomy VII}}, edited by
  \bibinfo{editor}{\bibfnamefont{W.~S.} \bibnamefont{{Holland}}}
  \bibnamefont{and}
  \bibinfo{editor}{\bibfnamefont{J.}~\bibnamefont{{Zmuidzinas}}}
  (\bibinfo{year}{2014}), vol. \bibinfo{volume}{9153} of
  \emph{\bibinfo{series}{Society of Photo-Optical Instrumentation Engineers
  (SPIE) Conference Series}}, p. \bibinfo{pages}{91531W}.

\bibitem[{\citenamefont{Dor{\'e} et~al.}(2018)\citenamefont{Dor{\'e}, Werner,
  Ashby, Bleem, Bock, Burt, Capak, Chang, Chaves-Montero, Chen
  et~al.}}]{SPHEREx-science-paper2018}
\bibinfo{author}{\bibfnamefont{O.}~\bibnamefont{Dor{\'e}}},
  \bibinfo{author}{\bibfnamefont{M.~W.} \bibnamefont{Werner}},
  \bibinfo{author}{\bibfnamefont{M.~L.} \bibnamefont{Ashby}},
  \bibinfo{author}{\bibfnamefont{L.~E.} \bibnamefont{Bleem}},
  \bibinfo{author}{\bibfnamefont{J.}~\bibnamefont{Bock}},
  \bibinfo{author}{\bibfnamefont{J.}~\bibnamefont{Burt}},
  \bibinfo{author}{\bibfnamefont{P.}~\bibnamefont{Capak}},
  \bibinfo{author}{\bibfnamefont{T.-C.} \bibnamefont{Chang}},
  \bibinfo{author}{\bibfnamefont{J.}~\bibnamefont{Chaves-Montero}},
  \bibinfo{author}{\bibfnamefont{C.~H.} \bibnamefont{Chen}},
  \bibnamefont{et~al.}, \bibinfo{journal}{arXiv preprint arXiv:1805.05489}
  (\bibinfo{year}{2018}).

\bibitem[{\citenamefont{Ade
  et~al.}(2020{\natexlab{a}})}]{CONCERTO-science-2020}
\bibinfo{author}{\bibfnamefont{P.}~\bibnamefont{Ade}} \bibnamefont{et~al.}
  (\bibinfo{collaboration}{CONCERTO}), \bibinfo{journal}{Astron. Astrophys.}
  \textbf{\bibinfo{volume}{642}}, \bibinfo{pages}{A60}
  (\bibinfo{year}{2020}{\natexlab{a}}), \eprint{2007.14246}.

\bibitem[{\citenamefont{Ade et~al.}(2020{\natexlab{b}})}]{EXCLAIM-2020}
\bibinfo{author}{\bibfnamefont{P.~A.~R.} \bibnamefont{Ade}}
  \bibnamefont{et~al.}, \bibinfo{journal}{J. Low Temp. Phys.}
  \textbf{\bibinfo{volume}{199}}, \bibinfo{pages}{1027}
  (\bibinfo{year}{2020}{\natexlab{b}}), \eprint{1912.07118}.

\bibitem[{\citenamefont{Aravena et~al.}(2022)\citenamefont{Aravena, Austermann,
  Basu, Battaglia, Beringue, Bertoldi, Bigiel, Bond, Breysse, Broughton
  et~al.}}]{CCAT-prime2021}
\bibinfo{author}{\bibfnamefont{M.}~\bibnamefont{Aravena}},
  \bibinfo{author}{\bibfnamefont{J.~E.} \bibnamefont{Austermann}},
  \bibinfo{author}{\bibfnamefont{K.}~\bibnamefont{Basu}},
  \bibinfo{author}{\bibfnamefont{N.}~\bibnamefont{Battaglia}},
  \bibinfo{author}{\bibfnamefont{B.}~\bibnamefont{Beringue}},
  \bibinfo{author}{\bibfnamefont{F.}~\bibnamefont{Bertoldi}},
  \bibinfo{author}{\bibfnamefont{F.}~\bibnamefont{Bigiel}},
  \bibinfo{author}{\bibfnamefont{J.~R.} \bibnamefont{Bond}},
  \bibinfo{author}{\bibfnamefont{P.~C.} \bibnamefont{Breysse}},
  \bibinfo{author}{\bibfnamefont{C.}~\bibnamefont{Broughton}},
  \bibnamefont{et~al.}, \bibinfo{journal}{The Astrophysical Journal Supplement
  Series} \textbf{\bibinfo{volume}{264}}, \bibinfo{pages}{7}
  (\bibinfo{year}{2022}).

\bibitem[{\citenamefont{Cleary et~al.}(2022)\citenamefont{Cleary, Borowska,
  Breysse, Catha, Chung, Church, Dickinson, Eriksen, Foss, Gundersen
  et~al.}}]{COmap-science-2021}
\bibinfo{author}{\bibfnamefont{K.~A.} \bibnamefont{Cleary}},
  \bibinfo{author}{\bibfnamefont{J.}~\bibnamefont{Borowska}},
  \bibinfo{author}{\bibfnamefont{P.~C.} \bibnamefont{Breysse}},
  \bibinfo{author}{\bibfnamefont{M.}~\bibnamefont{Catha}},
  \bibinfo{author}{\bibfnamefont{D.~T.} \bibnamefont{Chung}},
  \bibinfo{author}{\bibfnamefont{S.~E.} \bibnamefont{Church}},
  \bibinfo{author}{\bibfnamefont{C.}~\bibnamefont{Dickinson}},
  \bibinfo{author}{\bibfnamefont{H.~K.} \bibnamefont{Eriksen}},
  \bibinfo{author}{\bibfnamefont{M.~K.} \bibnamefont{Foss}},
  \bibinfo{author}{\bibfnamefont{J.~O.} \bibnamefont{Gundersen}},
  \bibnamefont{et~al.}, \bibinfo{journal}{The Astrophysical Journal}
  \textbf{\bibinfo{volume}{933}}, \bibinfo{pages}{182} (\bibinfo{year}{2022}).

\bibitem[{\citenamefont{Greve et~al.}(2014)\citenamefont{Greve, Leonidaki,
  Xilouris, Weiss, Zhang, van~der Werf, Aalto, Armus, Diaz-Santos, Evans
  et~al.}}]{Greve2014}
\bibinfo{author}{\bibfnamefont{T.}~\bibnamefont{Greve}},
  \bibinfo{author}{\bibfnamefont{I.}~\bibnamefont{Leonidaki}},
  \bibinfo{author}{\bibfnamefont{E.}~\bibnamefont{Xilouris}},
  \bibinfo{author}{\bibfnamefont{A.}~\bibnamefont{Weiss}},
  \bibinfo{author}{\bibfnamefont{Z.-Y.} \bibnamefont{Zhang}},
  \bibinfo{author}{\bibfnamefont{P.}~\bibnamefont{van~der Werf}},
  \bibinfo{author}{\bibfnamefont{S.}~\bibnamefont{Aalto}},
  \bibinfo{author}{\bibfnamefont{L.}~\bibnamefont{Armus}},
  \bibinfo{author}{\bibfnamefont{T.}~\bibnamefont{Diaz-Santos}},
  \bibinfo{author}{\bibfnamefont{A.}~\bibnamefont{Evans}},
  \bibnamefont{et~al.}, \bibinfo{journal}{arXiv preprint arXiv:1407.4400}
  (\bibinfo{year}{2014}).

\bibitem[{\citenamefont{Genzel et~al.}(2015)\citenamefont{Genzel, Tacconi,
  Lutz, Saintonge, Berta, Magnelli, Combes, Garc{\'\i}a-Burillo, Neri, Bolatto
  et~al.}}]{genzel2015combined}
\bibinfo{author}{\bibfnamefont{R.}~\bibnamefont{Genzel}},
  \bibinfo{author}{\bibfnamefont{L.}~\bibnamefont{Tacconi}},
  \bibinfo{author}{\bibfnamefont{D.}~\bibnamefont{Lutz}},
  \bibinfo{author}{\bibfnamefont{A.}~\bibnamefont{Saintonge}},
  \bibinfo{author}{\bibfnamefont{S.}~\bibnamefont{Berta}},
  \bibinfo{author}{\bibfnamefont{B.}~\bibnamefont{Magnelli}},
  \bibinfo{author}{\bibfnamefont{F.}~\bibnamefont{Combes}},
  \bibinfo{author}{\bibfnamefont{S.}~\bibnamefont{Garc{\'\i}a-Burillo}},
  \bibinfo{author}{\bibfnamefont{R.}~\bibnamefont{Neri}},
  \bibinfo{author}{\bibfnamefont{A.}~\bibnamefont{Bolatto}},
  \bibnamefont{et~al.}, \bibinfo{journal}{The Astrophysical Journal}
  \textbf{\bibinfo{volume}{800}}, \bibinfo{pages}{20} (\bibinfo{year}{2015}).

\bibitem[{\citenamefont{Schaerer et~al.}(2020)\citenamefont{Schaerer, Ginolfi,
  B{\'e}thermin, Fudamoto, Oesch, Le~F{\`e}vre, Faisst, Capak, Cassata,
  Silverman et~al.}}]{schaerer2020alpine}
\bibinfo{author}{\bibfnamefont{D.}~\bibnamefont{Schaerer}},
  \bibinfo{author}{\bibfnamefont{M.}~\bibnamefont{Ginolfi}},
  \bibinfo{author}{\bibfnamefont{M.}~\bibnamefont{B{\'e}thermin}},
  \bibinfo{author}{\bibfnamefont{Y.}~\bibnamefont{Fudamoto}},
  \bibinfo{author}{\bibfnamefont{P.}~\bibnamefont{Oesch}},
  \bibinfo{author}{\bibfnamefont{O.}~\bibnamefont{Le~F{\`e}vre}},
  \bibinfo{author}{\bibfnamefont{A.}~\bibnamefont{Faisst}},
  \bibinfo{author}{\bibfnamefont{P.}~\bibnamefont{Capak}},
  \bibinfo{author}{\bibfnamefont{P.}~\bibnamefont{Cassata}},
  \bibinfo{author}{\bibfnamefont{J.}~\bibnamefont{Silverman}},
  \bibnamefont{et~al.}, \bibinfo{journal}{Astronomy \& Astrophysics}
  \textbf{\bibinfo{volume}{643}}, \bibinfo{pages}{A3} (\bibinfo{year}{2020}).

\bibitem[{\citenamefont{Silva et~al.}(2021)\citenamefont{Silva, Kovetz,
  Keating, Dizgah, Bethermin, Breysse, Karkare, Bernal, and
  Delabrouille}}]{silva2021mapping}
\bibinfo{author}{\bibfnamefont{M.~B.} \bibnamefont{Silva}},
  \bibinfo{author}{\bibfnamefont{E.~D.} \bibnamefont{Kovetz}},
  \bibinfo{author}{\bibfnamefont{G.~K.} \bibnamefont{Keating}},
  \bibinfo{author}{\bibfnamefont{A.~M.} \bibnamefont{Dizgah}},
  \bibinfo{author}{\bibfnamefont{M.}~\bibnamefont{Bethermin}},
  \bibinfo{author}{\bibfnamefont{P.~C.} \bibnamefont{Breysse}},
  \bibinfo{author}{\bibfnamefont{K.}~\bibnamefont{Karkare}},
  \bibinfo{author}{\bibfnamefont{J.~L.} \bibnamefont{Bernal}},
  \bibnamefont{and}
  \bibinfo{author}{\bibfnamefont{J.}~\bibnamefont{Delabrouille}},
  \bibinfo{journal}{Experimental Astronomy} \textbf{\bibinfo{volume}{51}},
  \bibinfo{pages}{1593} (\bibinfo{year}{2021}).

\bibitem[{\citenamefont{Padmanabhan}(2021)}]{padmanabhan2021multi}
\bibinfo{author}{\bibfnamefont{H.}~\bibnamefont{Padmanabhan}},
  \bibinfo{journal}{International Journal of Modern Physics D}
  \textbf{\bibinfo{volume}{30}}, \bibinfo{pages}{2130009}
  (\bibinfo{year}{2021}).

\bibitem[{\citenamefont{Suginohara et~al.}(1998)\citenamefont{Suginohara,
  Suginohara, and Spergel}}]{Suginohara1998}
\bibinfo{author}{\bibfnamefont{M.}~\bibnamefont{Suginohara}},
  \bibinfo{author}{\bibfnamefont{T.}~\bibnamefont{Suginohara}},
  \bibnamefont{and} \bibinfo{author}{\bibfnamefont{D.}~\bibnamefont{Spergel}},
  \bibinfo{journal}{The Astrophysical Journal} \textbf{\bibinfo{volume}{512}},
  \bibinfo{pages}{547} (\bibinfo{year}{1998}).

\bibitem[{\citenamefont{Righi et~al.}(2008)\citenamefont{Righi,
  Hernandez-Monteagudo, and Sunyaev}}]{Righi2008b}
\bibinfo{author}{\bibfnamefont{M.}~\bibnamefont{Righi}},
  \bibinfo{author}{\bibfnamefont{C.}~\bibnamefont{Hernandez-Monteagudo}},
  \bibnamefont{and} \bibinfo{author}{\bibfnamefont{R.}~\bibnamefont{Sunyaev}},
  \bibinfo{journal}{Astron. Astrophys.} \textbf{\bibinfo{volume}{489}},
  \bibinfo{pages}{489} (\bibinfo{year}{2008}), \eprint{0805.2174}.

\bibitem[{\citenamefont{Carilli}(2011)}]{Carilli2011}
\bibinfo{author}{\bibfnamefont{C.}~\bibnamefont{Carilli}},
  \bibinfo{journal}{The Astrophysical Journal Letters}
  \textbf{\bibinfo{volume}{730}}, \bibinfo{pages}{L30} (\bibinfo{year}{2011}).

\bibitem[{\citenamefont{Fonseca et~al.}(2017)\citenamefont{Fonseca, Silva,
  Santos, and Cooray}}]{Fonseca2016}
\bibinfo{author}{\bibfnamefont{J.}~\bibnamefont{Fonseca}},
  \bibinfo{author}{\bibfnamefont{M.~B.} \bibnamefont{Silva}},
  \bibinfo{author}{\bibfnamefont{M.~G.} \bibnamefont{Santos}},
  \bibnamefont{and} \bibinfo{author}{\bibfnamefont{A.}~\bibnamefont{Cooray}},
  \bibinfo{journal}{Mon. Not. Roy. Astron. Soc.}
  \textbf{\bibinfo{volume}{464}}, \bibinfo{pages}{1948} (\bibinfo{year}{2017}).

\bibitem[{\citenamefont{Gong et~al.}(2017)\citenamefont{Gong, Cooray, Silva,
  Zemcov, Feng, Santos, Dore, and Chen}}]{Gong2017}
\bibinfo{author}{\bibfnamefont{Y.}~\bibnamefont{Gong}},
  \bibinfo{author}{\bibfnamefont{A.}~\bibnamefont{Cooray}},
  \bibinfo{author}{\bibfnamefont{M.~B.} \bibnamefont{Silva}},
  \bibinfo{author}{\bibfnamefont{M.}~\bibnamefont{Zemcov}},
  \bibinfo{author}{\bibfnamefont{C.}~\bibnamefont{Feng}},
  \bibinfo{author}{\bibfnamefont{M.~G.} \bibnamefont{Santos}},
  \bibinfo{author}{\bibfnamefont{O.}~\bibnamefont{Dore}}, \bibnamefont{and}
  \bibinfo{author}{\bibfnamefont{X.}~\bibnamefont{Chen}}, \bibinfo{journal}{The
  Astrophysical Journal} \textbf{\bibinfo{volume}{835}}, \bibinfo{pages}{273}
  (\bibinfo{year}{2017}).

\bibitem[{\citenamefont{Chung et~al.}(2020)\citenamefont{Chung, Viero, Church,
  and Wechsler}}]{Chung2018CII}
\bibinfo{author}{\bibfnamefont{D.~T.} \bibnamefont{Chung}},
  \bibinfo{author}{\bibfnamefont{M.~P.} \bibnamefont{Viero}},
  \bibinfo{author}{\bibfnamefont{S.~E.} \bibnamefont{Church}},
  \bibnamefont{and} \bibinfo{author}{\bibfnamefont{R.~H.}
  \bibnamefont{Wechsler}}, \bibinfo{journal}{Astrophys. J.}
  \textbf{\bibinfo{volume}{892}}, \bibinfo{pages}{51} (\bibinfo{year}{2020}),
  \eprint{1812.08135}.

\bibitem[{\citenamefont{Padmanabhan}(2018)}]{PadmanabhanCO}
\bibinfo{author}{\bibfnamefont{H.}~\bibnamefont{Padmanabhan}},
  \bibinfo{journal}{Mon. Not. Roy. Astron. Soc.}
  \textbf{\bibinfo{volume}{475}}, \bibinfo{pages}{1477} (\bibinfo{year}{2018}),
  \eprint{1706.01471}.

\bibitem[{\citenamefont{Padmanabhan}(2019)}]{Padmanabhan_CII}
\bibinfo{author}{\bibfnamefont{H.}~\bibnamefont{Padmanabhan}},
  \bibinfo{journal}{Mon. Not. Roy. Astron. Soc.}
  \textbf{\bibinfo{volume}{488}}, \bibinfo{pages}{3014} (\bibinfo{year}{2019}),
  \eprint{1811.01968}.

\bibitem[{\citenamefont{Dumitru et~al.}(2019)\citenamefont{Dumitru, Kulkarni,
  Lagache, and Haehnelt}}]{Dumitru2018}
\bibinfo{author}{\bibfnamefont{S.}~\bibnamefont{Dumitru}},
  \bibinfo{author}{\bibfnamefont{G.}~\bibnamefont{Kulkarni}},
  \bibinfo{author}{\bibfnamefont{G.}~\bibnamefont{Lagache}}, \bibnamefont{and}
  \bibinfo{author}{\bibfnamefont{M.~G.} \bibnamefont{Haehnelt}},
  \bibinfo{journal}{Mon. Not. Roy. Astron. Soc.}
  \textbf{\bibinfo{volume}{485}}, \bibinfo{pages}{3486} (\bibinfo{year}{2019}),
  \eprint{1802.04804}.

\bibitem[{\citenamefont{Chung et~al.}(2019)}]{Chung2018CO}
\bibinfo{author}{\bibfnamefont{D.~T.} \bibnamefont{Chung}} \bibnamefont{et~al.}
  (\bibinfo{collaboration}{COMAP}), \bibinfo{journal}{Astrophys. J.}
  \textbf{\bibinfo{volume}{872}}, \bibinfo{pages}{186} (\bibinfo{year}{2019}),
  \eprint{1809.04550}.

\bibitem[{\citenamefont{Kannan et~al.}(2022)\citenamefont{Kannan, Smith,
  Garaldi, Shen, Vogelsberger, Pakmor, Springel, and
  Hernquist}}]{Kannan:2021ucy}
\bibinfo{author}{\bibfnamefont{R.}~\bibnamefont{Kannan}},
  \bibinfo{author}{\bibfnamefont{A.}~\bibnamefont{Smith}},
  \bibinfo{author}{\bibfnamefont{E.}~\bibnamefont{Garaldi}},
  \bibinfo{author}{\bibfnamefont{X.}~\bibnamefont{Shen}},
  \bibinfo{author}{\bibfnamefont{M.}~\bibnamefont{Vogelsberger}},
  \bibinfo{author}{\bibfnamefont{R.}~\bibnamefont{Pakmor}},
  \bibinfo{author}{\bibfnamefont{V.}~\bibnamefont{Springel}}, \bibnamefont{and}
  \bibinfo{author}{\bibfnamefont{L.}~\bibnamefont{Hernquist}},
  \bibinfo{journal}{Mon. Not. Roy. Astron. Soc.}
  \textbf{\bibinfo{volume}{514}}, \bibinfo{pages}{3857} (\bibinfo{year}{2022}),
  \eprint{2111.02411}.

\bibitem[{\citenamefont{Murmu et~al.}(2021)\citenamefont{Murmu, Majumdar, and
  Datta}}]{Murmu:2021quo}
\bibinfo{author}{\bibfnamefont{C.~S.} \bibnamefont{Murmu}},
  \bibinfo{author}{\bibfnamefont{S.}~\bibnamefont{Majumdar}}, \bibnamefont{and}
  \bibinfo{author}{\bibfnamefont{K.~K.} \bibnamefont{Datta}},
  \bibinfo{journal}{Monthly Notices of the Royal Astronomical Society}
  \textbf{\bibinfo{volume}{507}}, \bibinfo{pages}{2500} (\bibinfo{year}{2021}).

\bibitem[{\citenamefont{Karoumpis et~al.}(2022)\citenamefont{Karoumpis,
  Magnelli, Romano-D{\'\i}az, Haslbauer, and Bertoldi}}]{Karoumpis2021}
\bibinfo{author}{\bibfnamefont{C.}~\bibnamefont{Karoumpis}},
  \bibinfo{author}{\bibfnamefont{B.}~\bibnamefont{Magnelli}},
  \bibinfo{author}{\bibfnamefont{E.}~\bibnamefont{Romano-D{\'\i}az}},
  \bibinfo{author}{\bibfnamefont{M.}~\bibnamefont{Haslbauer}},
  \bibnamefont{and} \bibinfo{author}{\bibfnamefont{F.}~\bibnamefont{Bertoldi}},
  \bibinfo{journal}{Astronomy \& Astrophysics} \textbf{\bibinfo{volume}{659}},
  \bibinfo{pages}{A12} (\bibinfo{year}{2022}).

\bibitem[{\citenamefont{Sun et~al.}(2023)\citenamefont{Sun, Mas-Ribas, Chang,
  Furlanetto, Mebane, Gonzalez, Parsons, and Trapp}}]{limfast2}
\bibinfo{author}{\bibfnamefont{G.}~\bibnamefont{Sun}},
  \bibinfo{author}{\bibfnamefont{L.}~\bibnamefont{Mas-Ribas}},
  \bibinfo{author}{\bibfnamefont{T.-C.} \bibnamefont{Chang}},
  \bibinfo{author}{\bibfnamefont{S.~R.} \bibnamefont{Furlanetto}},
  \bibinfo{author}{\bibfnamefont{R.~H.} \bibnamefont{Mebane}},
  \bibinfo{author}{\bibfnamefont{M.~O.} \bibnamefont{Gonzalez}},
  \bibinfo{author}{\bibfnamefont{J.}~\bibnamefont{Parsons}}, \bibnamefont{and}
  \bibinfo{author}{\bibfnamefont{A.}~\bibnamefont{Trapp}},
  \bibinfo{journal}{The Astrophysical Journal} \textbf{\bibinfo{volume}{950}},
  \bibinfo{pages}{40} (\bibinfo{year}{2023}).

\bibitem[{\citenamefont{Garcia et~al.}(2023)\citenamefont{Garcia, Narayanan,
  Popping, Anirudh, Sutherland, and Kaasinen}}]{garcia2023texttt}
\bibinfo{author}{\bibfnamefont{K.}~\bibnamefont{Garcia}},
  \bibinfo{author}{\bibfnamefont{D.}~\bibnamefont{Narayanan}},
  \bibinfo{author}{\bibfnamefont{G.}~\bibnamefont{Popping}},
  \bibinfo{author}{\bibfnamefont{R.}~\bibnamefont{Anirudh}},
  \bibinfo{author}{\bibfnamefont{S.}~\bibnamefont{Sutherland}},
  \bibnamefont{and} \bibinfo{author}{\bibfnamefont{M.}~\bibnamefont{Kaasinen}},
  \bibinfo{journal}{arXiv preprint arXiv:2311.01508}  (\bibinfo{year}{2023}).

\bibitem[{\citenamefont{Murmu et~al.}(2023)\citenamefont{Murmu, Olsen, Greve,
  Majumdar, Datta, Scott, Leung, Dav{\'e}, Popping, Ochoa
  et~al.}}]{Murmu:2021ljb}
\bibinfo{author}{\bibfnamefont{C.~S.} \bibnamefont{Murmu}},
  \bibinfo{author}{\bibfnamefont{K.~P.} \bibnamefont{Olsen}},
  \bibinfo{author}{\bibfnamefont{T.~R.} \bibnamefont{Greve}},
  \bibinfo{author}{\bibfnamefont{S.}~\bibnamefont{Majumdar}},
  \bibinfo{author}{\bibfnamefont{K.~K.} \bibnamefont{Datta}},
  \bibinfo{author}{\bibfnamefont{B.~R.} \bibnamefont{Scott}},
  \bibinfo{author}{\bibfnamefont{T.~D.} \bibnamefont{Leung}},
  \bibinfo{author}{\bibfnamefont{R.}~\bibnamefont{Dav{\'e}}},
  \bibinfo{author}{\bibfnamefont{G.}~\bibnamefont{Popping}},
  \bibinfo{author}{\bibfnamefont{R.~O.} \bibnamefont{Ochoa}},
  \bibnamefont{et~al.}, \bibinfo{journal}{Monthly Notices of the Royal
  Astronomical Society} \textbf{\bibinfo{volume}{518}}, \bibinfo{pages}{3074}
  (\bibinfo{year}{2023}).

\bibitem[{\citenamefont{Roy et~al.}(2023)\citenamefont{Roy,
  Valent{\'\i}n-Mart{\'\i}nez, Wang, Battaglia, and van Engelen}}]{limpy-roy}
\bibinfo{author}{\bibfnamefont{A.}~\bibnamefont{Roy}},
  \bibinfo{author}{\bibfnamefont{D.}~\bibnamefont{Valent{\'\i}n-Mart{\'\i}nez}},
  \bibinfo{author}{\bibfnamefont{K.}~\bibnamefont{Wang}},
  \bibinfo{author}{\bibfnamefont{N.}~\bibnamefont{Battaglia}},
  \bibnamefont{and} \bibinfo{author}{\bibfnamefont{A.}~\bibnamefont{van
  Engelen}}, \bibinfo{journal}{arXiv preprint arXiv:2304.06748}
  (\bibinfo{year}{2023}).

\bibitem[{\citenamefont{Puglisi et~al.}(2017)\citenamefont{Puglisi, Fabbian,
  and Baccigalupi}}]{puglisi:2017eqj}
\bibinfo{author}{\bibfnamefont{G.}~\bibnamefont{Puglisi}},
  \bibinfo{author}{\bibfnamefont{G.}~\bibnamefont{Fabbian}}, \bibnamefont{and}
  \bibinfo{author}{\bibfnamefont{C.}~\bibnamefont{Baccigalupi}},
  \bibinfo{journal}{Mon. Not. Roy. Astron. Soc.}
  \textbf{\bibinfo{volume}{469}}, \bibinfo{pages}{2982} (\bibinfo{year}{2017}),
  \eprint{1701.07856}.

\bibitem[{\citenamefont{Maniyar et~al.}(2023)\citenamefont{Maniyar, Gkogkou,
  Coulton, Li, Lagache, and Pullen}}]{maniyar2023extragalactic}
\bibinfo{author}{\bibfnamefont{A.~S.} \bibnamefont{Maniyar}},
  \bibinfo{author}{\bibfnamefont{A.}~\bibnamefont{Gkogkou}},
  \bibinfo{author}{\bibfnamefont{W.~R.} \bibnamefont{Coulton}},
  \bibinfo{author}{\bibfnamefont{Z.}~\bibnamefont{Li}},
  \bibinfo{author}{\bibfnamefont{G.}~\bibnamefont{Lagache}}, \bibnamefont{and}
  \bibinfo{author}{\bibfnamefont{A.~R.} \bibnamefont{Pullen}},
  \bibinfo{journal}{Physical Review D} \textbf{\bibinfo{volume}{107}},
  \bibinfo{pages}{123504} (\bibinfo{year}{2023}).

\bibitem[{\citenamefont{Ade et~al.}(2014{\natexlab{a}})\citenamefont{Ade,
  Aghanim, Alves, Armitage-Caplan, Arnaud, Ashdown, Atrio-Barandela, Aumont,
  Baccigalupi, Banday et~al.}}]{Planck-CO-2013}
\bibinfo{author}{\bibfnamefont{P.~A.} \bibnamefont{Ade}},
  \bibinfo{author}{\bibfnamefont{N.}~\bibnamefont{Aghanim}},
  \bibinfo{author}{\bibfnamefont{M.}~\bibnamefont{Alves}},
  \bibinfo{author}{\bibfnamefont{C.}~\bibnamefont{Armitage-Caplan}},
  \bibinfo{author}{\bibfnamefont{M.}~\bibnamefont{Arnaud}},
  \bibinfo{author}{\bibfnamefont{M.}~\bibnamefont{Ashdown}},
  \bibinfo{author}{\bibfnamefont{F.}~\bibnamefont{Atrio-Barandela}},
  \bibinfo{author}{\bibfnamefont{J.}~\bibnamefont{Aumont}},
  \bibinfo{author}{\bibfnamefont{C.}~\bibnamefont{Baccigalupi}},
  \bibinfo{author}{\bibfnamefont{A.}~\bibnamefont{Banday}},
  \bibnamefont{et~al.}, \bibinfo{journal}{Astronomy \& Astrophysics}
  \textbf{\bibinfo{volume}{571}}, \bibinfo{pages}{A13}
  (\bibinfo{year}{2014}{\natexlab{a}}).

\bibitem[{\citenamefont{Pullen et~al.}(2018)\citenamefont{Pullen, Serra, Chang,
  Dore, and Ho}}]{Pullen-CII-2018}
\bibinfo{author}{\bibfnamefont{A.~R.} \bibnamefont{Pullen}},
  \bibinfo{author}{\bibfnamefont{P.}~\bibnamefont{Serra}},
  \bibinfo{author}{\bibfnamefont{T.-C.} \bibnamefont{Chang}},
  \bibinfo{author}{\bibfnamefont{O.}~\bibnamefont{Dore}}, \bibnamefont{and}
  \bibinfo{author}{\bibfnamefont{S.}~\bibnamefont{Ho}}, \bibinfo{journal}{Mon.
  Not. Roy. Astron. Soc.} \textbf{\bibinfo{volume}{478}}, \bibinfo{pages}{1911}
  (\bibinfo{year}{2018}), \eprint{1707.06172}.

\bibitem[{\citenamefont{Dunne et~al.}(2024)\citenamefont{Dunne, Cleary,
  Breysse, Chung, Ihle, Bond, Eriksen, Gundersen, Keating, Kim
  et~al.}}]{Comap-stacking}
\bibinfo{author}{\bibfnamefont{D.~A.} \bibnamefont{Dunne}},
  \bibinfo{author}{\bibfnamefont{K.~A.} \bibnamefont{Cleary}},
  \bibinfo{author}{\bibfnamefont{P.~C.} \bibnamefont{Breysse}},
  \bibinfo{author}{\bibfnamefont{D.~T.} \bibnamefont{Chung}},
  \bibinfo{author}{\bibfnamefont{H.~T.} \bibnamefont{Ihle}},
  \bibinfo{author}{\bibfnamefont{J.~R.} \bibnamefont{Bond}},
  \bibinfo{author}{\bibfnamefont{H.~K.} \bibnamefont{Eriksen}},
  \bibinfo{author}{\bibfnamefont{J.~O.} \bibnamefont{Gundersen}},
  \bibinfo{author}{\bibfnamefont{L.~C.} \bibnamefont{Keating}},
  \bibinfo{author}{\bibfnamefont{J.}~\bibnamefont{Kim}}, \bibnamefont{et~al.},
  \bibinfo{journal}{The Astrophysical Journal} \textbf{\bibinfo{volume}{965}},
  \bibinfo{pages}{7} (\bibinfo{year}{2024}).

\bibitem[{\citenamefont{Greco et~al.}(2015)\citenamefont{Greco, Hill, Spergel,
  and Battaglia}}]{greco2015stacked}
\bibinfo{author}{\bibfnamefont{J.~P.} \bibnamefont{Greco}},
  \bibinfo{author}{\bibfnamefont{J.~C.} \bibnamefont{Hill}},
  \bibinfo{author}{\bibfnamefont{D.~N.} \bibnamefont{Spergel}},
  \bibnamefont{and}
  \bibinfo{author}{\bibfnamefont{N.}~\bibnamefont{Battaglia}},
  \bibinfo{journal}{The Astrophysical Journal} \textbf{\bibinfo{volume}{808}},
  \bibinfo{pages}{151} (\bibinfo{year}{2015}).

\bibitem[{\citenamefont{Aghanim et~al.}(2016)\citenamefont{Aghanim, Arnaud,
  Ashdown, Aumont, Baccigalupi, Banday, Barreiro, Bartlett, Bartolo, Battaner
  et~al.}}]{aghanim2016planck-tsz-stacking}
\bibinfo{author}{\bibfnamefont{N.}~\bibnamefont{Aghanim}},
  \bibinfo{author}{\bibfnamefont{M.}~\bibnamefont{Arnaud}},
  \bibinfo{author}{\bibfnamefont{M.}~\bibnamefont{Ashdown}},
  \bibinfo{author}{\bibfnamefont{J.}~\bibnamefont{Aumont}},
  \bibinfo{author}{\bibfnamefont{C.}~\bibnamefont{Baccigalupi}},
  \bibinfo{author}{\bibfnamefont{A.}~\bibnamefont{Banday}},
  \bibinfo{author}{\bibfnamefont{R.}~\bibnamefont{Barreiro}},
  \bibinfo{author}{\bibfnamefont{J.}~\bibnamefont{Bartlett}},
  \bibinfo{author}{\bibfnamefont{N.}~\bibnamefont{Bartolo}},
  \bibinfo{author}{\bibfnamefont{E.}~\bibnamefont{Battaner}},
  \bibnamefont{et~al.}, \bibinfo{journal}{Astronomy \& Astrophysics}
  \textbf{\bibinfo{volume}{594}}, \bibinfo{pages}{A22} (\bibinfo{year}{2016}).

\bibitem[{\citenamefont{Hill et~al.}(2018)\citenamefont{Hill, Baxter, Lidz,
  Greco, and Jain}}]{hill2018two-stacking}
\bibinfo{author}{\bibfnamefont{J.~C.} \bibnamefont{Hill}},
  \bibinfo{author}{\bibfnamefont{E.~J.} \bibnamefont{Baxter}},
  \bibinfo{author}{\bibfnamefont{A.}~\bibnamefont{Lidz}},
  \bibinfo{author}{\bibfnamefont{J.~P.} \bibnamefont{Greco}}, \bibnamefont{and}
  \bibinfo{author}{\bibfnamefont{B.}~\bibnamefont{Jain}},
  \bibinfo{journal}{Physical Review D} \textbf{\bibinfo{volume}{97}},
  \bibinfo{pages}{083501} (\bibinfo{year}{2018}).

\bibitem[{\citenamefont{Hall et~al.}(2019)\citenamefont{Hall, Zakamska,
  Addison, Battaglia, Crichton, Devlin, Dunkley, Gralla, Hill, Hilton
  et~al.}}]{Hall-tsz-2019}
\bibinfo{author}{\bibfnamefont{K.~R.} \bibnamefont{Hall}},
  \bibinfo{author}{\bibfnamefont{N.~L.} \bibnamefont{Zakamska}},
  \bibinfo{author}{\bibfnamefont{G.~E.} \bibnamefont{Addison}},
  \bibinfo{author}{\bibfnamefont{N.}~\bibnamefont{Battaglia}},
  \bibinfo{author}{\bibfnamefont{D.}~\bibnamefont{Crichton}},
  \bibinfo{author}{\bibfnamefont{M.}~\bibnamefont{Devlin}},
  \bibinfo{author}{\bibfnamefont{J.}~\bibnamefont{Dunkley}},
  \bibinfo{author}{\bibfnamefont{M.}~\bibnamefont{Gralla}},
  \bibinfo{author}{\bibfnamefont{J.~C.} \bibnamefont{Hill}},
  \bibinfo{author}{\bibfnamefont{M.}~\bibnamefont{Hilton}},
  \bibnamefont{et~al.}, \bibinfo{journal}{Monthly Notices of the Royal
  Astronomical Society} \textbf{\bibinfo{volume}{490}}, \bibinfo{pages}{2315}
  (\bibinfo{year}{2019}).

\bibitem[{\citenamefont{Dawson et~al.}(2016)\citenamefont{Dawson, Kneib,
  Percival, Alam, Albareti, Anderson, Armengaud, Aubourg, Bailey, Bautista
  et~al.}}]{eboss-technical}
\bibinfo{author}{\bibfnamefont{K.~S.} \bibnamefont{Dawson}},
  \bibinfo{author}{\bibfnamefont{J.-P.} \bibnamefont{Kneib}},
  \bibinfo{author}{\bibfnamefont{W.~J.} \bibnamefont{Percival}},
  \bibinfo{author}{\bibfnamefont{S.}~\bibnamefont{Alam}},
  \bibinfo{author}{\bibfnamefont{F.~D.} \bibnamefont{Albareti}},
  \bibinfo{author}{\bibfnamefont{S.~F.} \bibnamefont{Anderson}},
  \bibinfo{author}{\bibfnamefont{E.}~\bibnamefont{Armengaud}},
  \bibinfo{author}{\bibfnamefont{{\'E}.}~\bibnamefont{Aubourg}},
  \bibinfo{author}{\bibfnamefont{S.}~\bibnamefont{Bailey}},
  \bibinfo{author}{\bibfnamefont{J.~E.} \bibnamefont{Bautista}},
  \bibnamefont{et~al.}, \bibinfo{journal}{The Astronomical Journal}
  \textbf{\bibinfo{volume}{151}}, \bibinfo{pages}{44} (\bibinfo{year}{2016}).

\bibitem[{\citenamefont{Blanton et~al.}(2017)\citenamefont{Blanton, Bershady,
  Abolfathi, Albareti, Prieto, Almeida, Alonso-Garc{\'\i}a, Anders, Anderson,
  Andrews et~al.}}]{eboss-overview}
\bibinfo{author}{\bibfnamefont{M.~R.} \bibnamefont{Blanton}},
  \bibinfo{author}{\bibfnamefont{M.~A.} \bibnamefont{Bershady}},
  \bibinfo{author}{\bibfnamefont{B.}~\bibnamefont{Abolfathi}},
  \bibinfo{author}{\bibfnamefont{F.~D.} \bibnamefont{Albareti}},
  \bibinfo{author}{\bibfnamefont{C.~A.} \bibnamefont{Prieto}},
  \bibinfo{author}{\bibfnamefont{A.}~\bibnamefont{Almeida}},
  \bibinfo{author}{\bibfnamefont{J.}~\bibnamefont{Alonso-Garc{\'\i}a}},
  \bibinfo{author}{\bibfnamefont{F.}~\bibnamefont{Anders}},
  \bibinfo{author}{\bibfnamefont{S.~F.} \bibnamefont{Anderson}},
  \bibinfo{author}{\bibfnamefont{B.}~\bibnamefont{Andrews}},
  \bibnamefont{et~al.}, \bibinfo{journal}{The Astronomical Journal}
  \textbf{\bibinfo{volume}{154}}, \bibinfo{pages}{28} (\bibinfo{year}{2017}).

\bibitem[{\citenamefont{Aghanim et~al.}(2020)\citenamefont{Aghanim, Akrami,
  Ashdown, Aumont, Baccigalupi, Ballardini, Banday, Barreiro, Bartolo, Basak
  et~al.}}]{P18:main}
\bibinfo{author}{\bibfnamefont{N.}~\bibnamefont{Aghanim}},
  \bibinfo{author}{\bibfnamefont{Y.}~\bibnamefont{Akrami}},
  \bibinfo{author}{\bibfnamefont{M.}~\bibnamefont{Ashdown}},
  \bibinfo{author}{\bibfnamefont{J.}~\bibnamefont{Aumont}},
  \bibinfo{author}{\bibfnamefont{C.}~\bibnamefont{Baccigalupi}},
  \bibinfo{author}{\bibfnamefont{M.}~\bibnamefont{Ballardini}},
  \bibinfo{author}{\bibfnamefont{A.~J.} \bibnamefont{Banday}},
  \bibinfo{author}{\bibfnamefont{R.}~\bibnamefont{Barreiro}},
  \bibinfo{author}{\bibfnamefont{N.}~\bibnamefont{Bartolo}},
  \bibinfo{author}{\bibfnamefont{S.}~\bibnamefont{Basak}},
  \bibnamefont{et~al.}, \bibinfo{journal}{Astronomy \& Astrophysics}
  \textbf{\bibinfo{volume}{641}}, \bibinfo{pages}{A6} (\bibinfo{year}{2020}).

\bibitem[{\citenamefont{Ade et~al.}(2014{\natexlab{b}})\citenamefont{Ade,
  Aghanim, Armitage-Caplan, Arnaud, Ashdown, Atrio-Barandela, Aumont,
  Baccigalupi, Banday, Barreiro et~al.}}]{Planck-HFI-spectral-response}
\bibinfo{author}{\bibfnamefont{P.~A.} \bibnamefont{Ade}},
  \bibinfo{author}{\bibfnamefont{N.}~\bibnamefont{Aghanim}},
  \bibinfo{author}{\bibfnamefont{C.}~\bibnamefont{Armitage-Caplan}},
  \bibinfo{author}{\bibfnamefont{M.}~\bibnamefont{Arnaud}},
  \bibinfo{author}{\bibfnamefont{M.}~\bibnamefont{Ashdown}},
  \bibinfo{author}{\bibfnamefont{F.}~\bibnamefont{Atrio-Barandela}},
  \bibinfo{author}{\bibfnamefont{J.}~\bibnamefont{Aumont}},
  \bibinfo{author}{\bibfnamefont{C.}~\bibnamefont{Baccigalupi}},
  \bibinfo{author}{\bibfnamefont{A.}~\bibnamefont{Banday}},
  \bibinfo{author}{\bibfnamefont{R.}~\bibnamefont{Barreiro}},
  \bibnamefont{et~al.}, \bibinfo{journal}{Astronomy \& Astrophysics}
  \textbf{\bibinfo{volume}{571}}, \bibinfo{pages}{A9}
  (\bibinfo{year}{2014}{\natexlab{b}}).

\bibitem[{\citenamefont{Gorski et~al.}(2005)\citenamefont{Gorski, Hivon,
  Banday, Wandelt, Hansen, Reinecke, and Bartelmann}}]{healpix}
\bibinfo{author}{\bibfnamefont{K.~M.} \bibnamefont{Gorski}},
  \bibinfo{author}{\bibfnamefont{E.}~\bibnamefont{Hivon}},
  \bibinfo{author}{\bibfnamefont{A.~J.} \bibnamefont{Banday}},
  \bibinfo{author}{\bibfnamefont{B.~D.} \bibnamefont{Wandelt}},
  \bibinfo{author}{\bibfnamefont{F.~K.} \bibnamefont{Hansen}},
  \bibinfo{author}{\bibfnamefont{M.}~\bibnamefont{Reinecke}}, \bibnamefont{and}
  \bibinfo{author}{\bibfnamefont{M.}~\bibnamefont{Bartelmann}},
  \bibinfo{journal}{The Astrophysical Journal} \textbf{\bibinfo{volume}{622}},
  \bibinfo{pages}{759} (\bibinfo{year}{2005}).

\bibitem[{\citenamefont{Price-Whelan et~al.}(2022)\citenamefont{Price-Whelan,
  Lim, Earl, Starkman, Bradley, Shupe, Patil, Corrales, Brasseur, N{\"o}the
  et~al.}}]{Astropy-5}
\bibinfo{author}{\bibfnamefont{A.~M.} \bibnamefont{Price-Whelan}},
  \bibinfo{author}{\bibfnamefont{P.~L.} \bibnamefont{Lim}},
  \bibinfo{author}{\bibfnamefont{N.}~\bibnamefont{Earl}},
  \bibinfo{author}{\bibfnamefont{N.}~\bibnamefont{Starkman}},
  \bibinfo{author}{\bibfnamefont{L.}~\bibnamefont{Bradley}},
  \bibinfo{author}{\bibfnamefont{D.~L.} \bibnamefont{Shupe}},
  \bibinfo{author}{\bibfnamefont{A.~A.} \bibnamefont{Patil}},
  \bibinfo{author}{\bibfnamefont{L.}~\bibnamefont{Corrales}},
  \bibinfo{author}{\bibfnamefont{C.}~\bibnamefont{Brasseur}},
  \bibinfo{author}{\bibfnamefont{M.}~\bibnamefont{N{\"o}the}},
  \bibnamefont{et~al.}, \bibinfo{journal}{The Astrophysical Journal}
  \textbf{\bibinfo{volume}{935}}, \bibinfo{pages}{167} (\bibinfo{year}{2022}).

\bibitem[{\citenamefont{Foreman-Mackey
  et~al.}(2013)\citenamefont{Foreman-Mackey, Hogg, Lang, and Goodman}}]{emcee}
\bibinfo{author}{\bibfnamefont{D.}~\bibnamefont{Foreman-Mackey}},
  \bibinfo{author}{\bibfnamefont{D.~W.} \bibnamefont{Hogg}},
  \bibinfo{author}{\bibfnamefont{D.}~\bibnamefont{Lang}}, \bibnamefont{and}
  \bibinfo{author}{\bibfnamefont{J.}~\bibnamefont{Goodman}},
  \bibinfo{journal}{Publications of the Astronomical Society of the Pacific}
  \textbf{\bibinfo{volume}{125}}, \bibinfo{pages}{306} (\bibinfo{year}{2013}).

\bibitem[{\citenamefont{Visbal and Loeb}(2010)}]{Visbal2010}
\bibinfo{author}{\bibfnamefont{E.}~\bibnamefont{Visbal}} \bibnamefont{and}
  \bibinfo{author}{\bibfnamefont{A.}~\bibnamefont{Loeb}},
  \bibinfo{journal}{Journal of Cosmology and Astroparticle Physics}
  \textbf{\bibinfo{volume}{2010}}, \bibinfo{pages}{016} (\bibinfo{year}{2010}).

\bibitem[{\citenamefont{Kamenetzky et~al.}(2016)\citenamefont{Kamenetzky,
  Rangwala, Glenn, Maloney, and Conley}}]{Kamenetzky2016}
\bibinfo{author}{\bibfnamefont{J.}~\bibnamefont{Kamenetzky}},
  \bibinfo{author}{\bibfnamefont{N.}~\bibnamefont{Rangwala}},
  \bibinfo{author}{\bibfnamefont{J.}~\bibnamefont{Glenn}},
  \bibinfo{author}{\bibfnamefont{P.}~\bibnamefont{Maloney}}, \bibnamefont{and}
  \bibinfo{author}{\bibfnamefont{A.}~\bibnamefont{Conley}},
  \bibinfo{journal}{The Astrophysical Journal} \textbf{\bibinfo{volume}{829}},
  \bibinfo{pages}{93} (\bibinfo{year}{2016}).

\bibitem[{\citenamefont{Nelson et~al.}(2018)}]{Nelson:2018uso}
\bibinfo{author}{\bibfnamefont{D.}~\bibnamefont{Nelson}} \bibnamefont{et~al.}
  (\bibinfo{year}{2018}), \eprint{1812.05609}.

\bibitem[{\citenamefont{Pillepich et~al.}(2018)}]{TNG-gal}
\bibinfo{author}{\bibfnamefont{A.}~\bibnamefont{Pillepich}}
  \bibnamefont{et~al.}, \bibinfo{journal}{Mon. Not. Roy. Astron. Soc.}
  \textbf{\bibinfo{volume}{475}}, \bibinfo{pages}{648} (\bibinfo{year}{2018}),
  \eprint{1707.03406}.

\bibitem[{\citenamefont{Springel et~al.}(2018)}]{TNG-gen}
\bibinfo{author}{\bibfnamefont{V.}~\bibnamefont{Springel}}
  \bibnamefont{et~al.}, \bibinfo{journal}{Mon. Not. Roy. Astron. Soc.}
  \textbf{\bibinfo{volume}{475}}, \bibinfo{pages}{676} (\bibinfo{year}{2018}),
  \eprint{1707.03397}.

\bibitem[{\citenamefont{Silva et~al.}(2015)\citenamefont{Silva, Santos, Cooray,
  and Gong}}]{Silva:2015}
\bibinfo{author}{\bibfnamefont{M.~B.} \bibnamefont{Silva}},
  \bibinfo{author}{\bibfnamefont{M.~G.} \bibnamefont{Santos}},
  \bibinfo{author}{\bibfnamefont{A.}~\bibnamefont{Cooray}}, \bibnamefont{and}
  \bibinfo{author}{\bibfnamefont{Y.}~\bibnamefont{Gong}},
  \bibinfo{journal}{Astrophys. J.} \textbf{\bibinfo{volume}{806}},
  \bibinfo{pages}{209} (\bibinfo{year}{2015}), \eprint{1410.4808}.

\bibitem[{\citenamefont{{Behroozi} et~al.}(2019)\citenamefont{{Behroozi},
  {Wechsler}, {Hearin}, and {Conroy}}}]{Behroozi2019}
\bibinfo{author}{\bibfnamefont{P.}~\bibnamefont{{Behroozi}}},
  \bibinfo{author}{\bibfnamefont{R.~H.} \bibnamefont{{Wechsler}}},
  \bibinfo{author}{\bibfnamefont{A.~P.} \bibnamefont{{Hearin}}},
  \bibnamefont{and} \bibinfo{author}{\bibfnamefont{C.}~\bibnamefont{{Conroy}}},
  \bibinfo{journal}{Mon. Not. Roy. Astron. Soc.}
  \textbf{\bibinfo{volume}{488}}, \bibinfo{pages}{3143} (\bibinfo{year}{2019}),
  \eprint{1806.07893}.

\bibitem[{\citenamefont{Breysse et~al.}(2022)\citenamefont{Breysse, Yang,
  Somerville, Pullen, Popping, and Maniyar}}]{Breysse2022-co}
\bibinfo{author}{\bibfnamefont{P.~C.} \bibnamefont{Breysse}},
  \bibinfo{author}{\bibfnamefont{S.}~\bibnamefont{Yang}},
  \bibinfo{author}{\bibfnamefont{R.~S.} \bibnamefont{Somerville}},
  \bibinfo{author}{\bibfnamefont{A.~R.} \bibnamefont{Pullen}},
  \bibinfo{author}{\bibfnamefont{G.}~\bibnamefont{Popping}}, \bibnamefont{and}
  \bibinfo{author}{\bibfnamefont{A.~S.} \bibnamefont{Maniyar}},
  \bibinfo{journal}{The Astrophysical Journal} \textbf{\bibinfo{volume}{929}},
  \bibinfo{pages}{30} (\bibinfo{year}{2022}).

\bibitem[{\citenamefont{Horlaville et~al.}(2024)\citenamefont{Horlaville,
  Chung, Bond, and Liang}}]{Horlaville2024}
\bibinfo{author}{\bibfnamefont{P.}~\bibnamefont{Horlaville}},
  \bibinfo{author}{\bibfnamefont{D.~T.} \bibnamefont{Chung}},
  \bibinfo{author}{\bibfnamefont{J.~R.} \bibnamefont{Bond}}, \bibnamefont{and}
  \bibinfo{author}{\bibfnamefont{L.}~\bibnamefont{Liang}},
  \bibinfo{journal}{Monthly Notices of the Royal Astronomical Society}
  \textbf{\bibinfo{volume}{531}}, \bibinfo{pages}{2958} (\bibinfo{year}{2024}).

\bibitem[{\citenamefont{{Kokron} et~al.}(2024)\citenamefont{{Kokron}, {Bernal},
  and {Dunkley}}}]{Kokron24}
\bibinfo{author}{\bibfnamefont{N.}~\bibnamefont{{Kokron}}},
  \bibinfo{author}{\bibfnamefont{J.~L.} \bibnamefont{{Bernal}}},
  \bibnamefont{and}
  \bibinfo{author}{\bibfnamefont{J.}~\bibnamefont{{Dunkley}}},
  \bibinfo{journal}{arXiv e-prints} \bibinfo{eid}{arXiv:2405.20369}
  (\bibinfo{year}{2024}), \eprint{2405.20369}.

\end{thebibliography}

\clearpage
\appendix
\section{Characterization of residual foreground}

We present a methodology for confirming the correlation between CO(3-2) emissions and galaxies within the frequency bandwidth observed by Planck. Specifically, we performed checks and tests to ensure that the CO(3-2) signal originates from galaxies corresponding to Planck's frequency range in the presence of residual foreground on the map. We emphasize that within the observational frequency range of 109.6–115.4 GHz, where the Planck CO(1-0) map has been constructed, there should not be correlated CO(3-2) emissions from galaxies in the redshift bin of 0.49 to 0.57. To verify this, we employ a stacking technique across galaxies grouped into four distinct redshift bins, aligning with the Planck CO(1-0) map. This approach enables us to effectively characterize and account for residual foreground contamination present in the map at 115 GHz.

\begin{figure*}[h]
\includegraphics[width=0.99\textwidth]{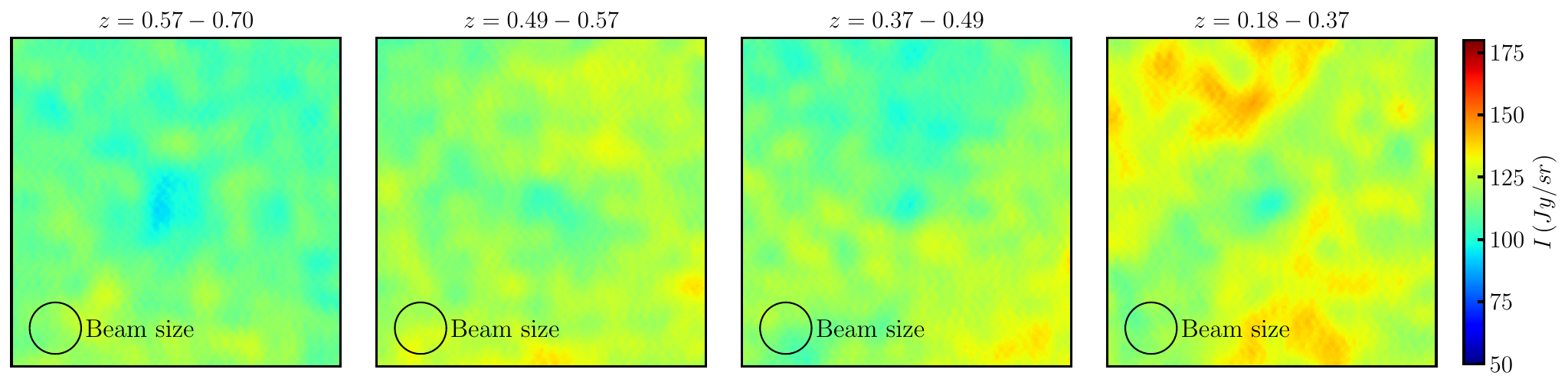} 
\caption{
The left-over contaminants in the map within the observational frequency bandwidth of 109.6 – 115.4 GHz. The absence of prominent structure at the center of the stacked images signifies that there is no CO(3-2) contribution from galaxies in the redshift bin 0.49 to 0.57, and also that the residual CIB contributions are smaller than those for the observational frequency range of 219.3 – 230.8 GHz. The assumed FWHM of the map's beam is $9'.65$ \cite{Planck-CO-2013}.}
\label{fig:sub_CIB}
\end{figure*}

\begin{figure*}[h]
\includegraphics[width=0.45\textwidth]{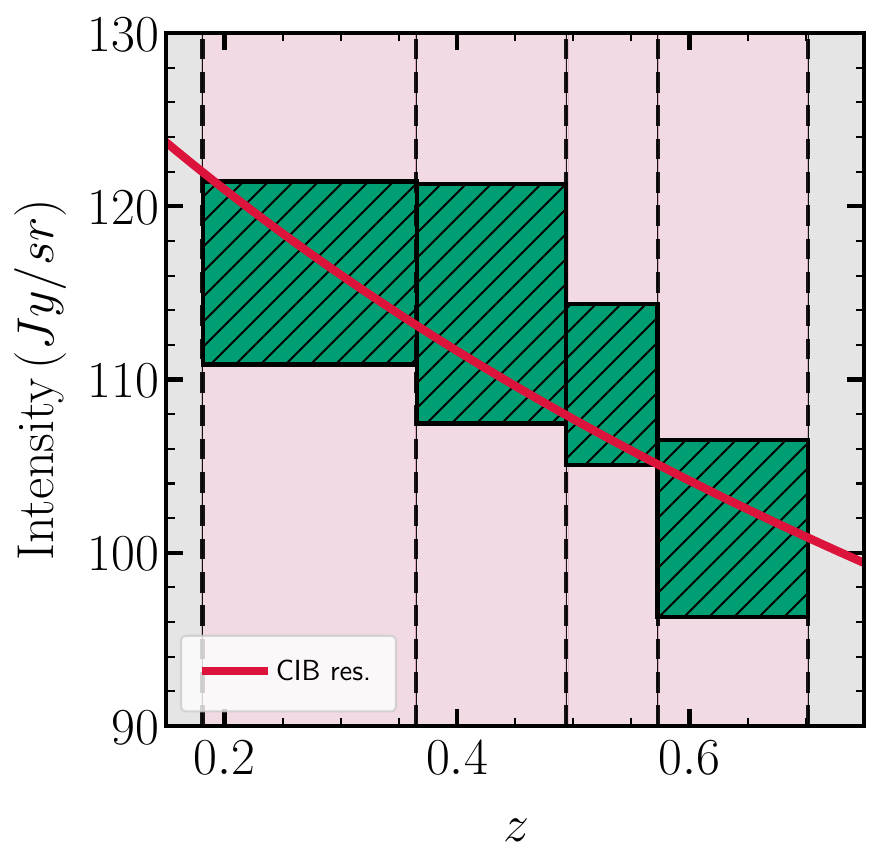} 
\caption{Mean intensities of stacked eBOSS galaxies overlaid onto Planck's CO(1-0) map. The boxes represent intensity variations, with box length indicating intensity uncertainty, and box width depicting the range of redshift bins. Additionally, the figure displays the best-fit power-law model for the residual foreground, aligning with the measured intensities.}
\label{fig:CIB-res-fit-CO10}
\end{figure*}

We show the stacking images at four redshifts in Figure \ref{fig:sub_CIB}, encompassing residual foreground emissions at 115 GHz. As CO emission for all J-level transitions should not be present at any of these four redshifts, we expect these images to capture the information of the residual foreground. As previously mentioned, we modeled the residual foreground as a power-law with two free parameters, $I^0_{\rm res}$ and $\alpha$. We aim to check the validity of this model by estimating the $\chi^2$ and PTE for the 115 GHz channel. The average intensity changes from $116.2 \pm 5.3$ Jy/sr to $101.3 \pm 4.6$ Jy/sr from the redshift bins 0.18--0.37 to 0.57--0.70. We show the distribution of mean intensity with redshift in Figure \ref{fig:CIB-res-fit-CO10}, which displays a power-law trend. We estimate the best-fit values of $I_{\rm res}^{0}$ and $\alpha$ to be $133^{+14}_{-12}\,(Jy/sr)$ and $0.52^{ +0.25}_{-0.24}$, respectively. The $\chi^2$ value is 0.55 with 2 degrees of freedom, and the PTE is 0.76. This test allows us to assume a power-law-like residual foreground model for the CO(2-1) map constructed at 230\,GHz to probe the CO(3-2) signal from the galaxies at $z\sim 0.5$.

\end{document}